\documentclass[12pt]{iopart}

\usepackage{graphicx} 
\begin{document}

\title[Non-Markovianity, entropy production, and Jarzynski equality]{Non-Markovianity, entropy production, and Jarzynski equality}

\author{Jackes Martins}
\address{Departamento de F\'{i}sica, Pontif\'{i}cia Universidade Cat\'{o}lica, 22452-970, Rio de Janeiro, Brazil}

\author{Lucianno Defaveri}
\address{Departamento de F\'{i}sica, Pontif\'{i}cia Universidade Cat\'{o}lica, 22452-970, Rio de Janeiro, Brazil}

\author{Diogo O. Soares-Pinto}
\address{ Instituto de F\'{i}sica de S\~{a}o Carlos, Universidade de S\~{a}o Paulo CP 369, 13560-970, S\~{a}o Carlos, S\~{a}o Paulo, Brazil}

\author{S\'{\i}lvio M. Duarte Queir\'{o}s}
\address{Centro Brasileiro de Pesquisas F\'{\i}sicas,
            Rua Dr Xavier Sigaud, 150, 22290-180 Rio de Janeiro -- RJ, Brazil, and National Institute of Science and Technology for Complex Systems}

\author{Welles A.M. Morgado}
\address{Departamento de F\'{i}sica, Pontif\'{i}cia Universidade Cat\'{o}lica, 22452-970, Rio de Janeiro, Brazil, National Institute of Science and Technology for Complex Systems}
\ead{welles@puc-rio.br}
\vspace{10pt}
\begin{indented}
\item[]
\end{indented}

\begin{abstract}
We explore the role a non-Markovian memory kernel plays on information exchange and entropy production in the context of a external work protocol. The Jarzynski Equality is shown to hold for both the harmonic and the non-harmonic models. We observe the memory function acts as an information pump, recovering part of the information lost to the thermal reservoir as a consequence of the non-equilibrium work protocol. The pumping action occurs for both the harmonic and non-harmonic cases. Unexpectedly, we found that the harmonic model does not produce entropy, regardless of the work protocol. The presence of even a small amount of non-linearity recovers the  more normal entropy producing behavior, for out-of-equilibrium protocols.  
\end{abstract}

%
%
%
%
%

\section{Introduction}
Fluctuation theorems, for entropy and work, provide us with an important, and sometimes unexpected, insight into the inner workings of non-equilibrium driven systems~\cite{Gallavotti1995,Evans1993,Evans1994,Jarzynski1997a,Crooks2000,Kurchan1998,Seifert2012,Jarzynski2011}. In the case of work, fluctuation theorems state the dissipated work obeys an exact, and restrictive, relation~\cite{Jarzynski1997a,Mazonka1999,Subasi2013}. These systems -- at first in an equilibrium state -- are then perturbed in a reproducible way, by means of a time protocol on some interaction variable. The validity of the Jarzynski relation has been shown in the case of a non-Markovian description of the  system~\cite{Speck2007}, characterized by a memory function responsible for the dissipation of energy~\cite{Zwanzig1961,Mori1965}. As we shall see, the memory function may act as a information backflow channel between system and thermal reservoir. Understanding the flow of information is important and may have many applications. Herein, we shall approach the question by studying, in detail, a non-linear non-Markovian Brownian particle model under the action of an external protocol. 

We approach the problem of information flow via the study of the interaction of a small system with a thermal bath. The system is also coupled to an external system by means of an external variable (piston position) that changes in time according to a predefined protocol. The present work has  similarities to an older model that was solved exactly~\cite{Morgado2010}, where the Jarzynski equality~\cite{Jarzynski1997} was verified for the harmonically bound particle under the action of white noise. Despite its somewhat simplicity, Langevin systems are quite useful and reproduce behavior found in more complex systems, such as Fluctuation Theorems~\cite{Kurchan1998}.

In the present work, we generalize the friction coefficient to include a memory kernel, leading to a non-Markovian behavior. We also generalize the potential to a non-harmonic interaction. The memory kernel expresses  exchange of information through time, which is at the core of non-Markovian behavior. Whereas harmonic systems are possible to treat mathematically, they present weaknesses in terms of their oversimplified physics. For instance, harmonic potentials do not couple well with higher order noise cumulants~\cite{Morgado2012b, Morgado2016} and efficient, but purely harmonic, machines cannot be build~\cite{Defaveri2017, Defaveri2018}. 

We shall use the Shannon out-of-equilibrium entropy in order to define the instantaneous entropy for the system~\cite{Parrondo2015,Esposito2011,Esposito2010a}. The  out-of-equilibrium entropy is a useful quantity for understanding the interplay of the information flow and the heat and work exchanges for the system. The piston (work) protocol acts as a control knob, allowing us to tune the exchanges of entropy and work with the environment. The entropic budget, production and exchanges between system and reservoir, show very intriguing properties such as: for the purely harmonic model the piston protocol does not produce entropy; there is constant backflow of information between system and reservoir.  By moving away from equilibrium limitations we hope to be able to understand the dynamics of information flow and control in these simple systems.

Indeed, one of the points we are most interested in is the relationship between the  Jarzynski equality, which can be seen as a restriction upon the non-equilibrium behavior of the system, and the memory kernel, which governs the flow of information to and from the system. The Jarzynski equality~\cite{Jarzynski1997,Jarzynski2007} acts as a strong constraint upon the system. Since it must be obeyed in all circumstances, it sets  the gauge for analytical and numerical models. It enhances the importance of very rare improbable events. In particular, there are ``free-lunch'' states, associated with energy being taken from the reservoir ``for free'', as we see below. In order to get insight into it, let us rewrite Eq.~\ref{JE} in terms of the dissipated work $W_{d} = W - \Delta F$:
\[
\left<e^{-\beta\,W_{d}}\right> = 1.
\]
Since the result above is protocol independent, we can choose a protocol that typically generates very far from equilibrium states (such as compressing a gas almost instantaneously), yielding very large values for the dissipated work. Typically 
\[
\beta\,\left<W_{d}\right> \gg 1 \Rightarrow 0 <e^{-\beta\,W_{d}} \ll 1.
\]

With very high probability, every time the experiment is repeated we get $e^{-\beta W_{d}}\ll 1$. So, how does the Jarzynski equality comes out true than? It is because some very improbable cases arise, with $W_{d}<0 \Rightarrow e^{-\beta\,W_{d}} > 1$, and take the average of the exponential back to 1. We shall call these  rare event states as Free-Lunches (FL)~\cite{Pellegrini2019}. We shall illustrate below a possible free-lunch for a far from equilibrium protocol.  We should keep in mind that a FL is the outcome of the initial condition and the external protocol applied to it.

Other entropic effects can be studied in the context of the Jarzynski equality. In particular, the non-Markovian Brownian model allows for the exchange of information along time due to the presence of the memory kernel. The excitation of slow hydrodynamic modes can generate retarded kernel functions that allow for the Brownian particle in the present to interact with its own state in the past. An interesting question arises: can it recover (partially at least) the information it lost, via dissipation, to the thermal bath earlier? We shall attempt to shed light on this topic. Also, we shall try and understand some of the roles played by the non-linearities on the production and transfer of entropy for these systems. 

In section II, we define the non-Markovian model its procedure and illustrate with an example of a so called ``free-lunch''. In section III, we solve the harmonic non-Markovian model exactly. In section IV, we study the properties of the entropy flux for the model. We describe the entropy oscillations and show that a harmonic model does not produce entropy. In section V, we exhibit the numerical results for the non-linear non-Markovian model. In section VI, we briefly discuss the results herein.

%
\section{The non-Markovian model}
\subsection{Model}
We shall study the Jarzynski equality by means of a massive Brownian particle under the action of external driving force and by a combination of harmonic and quartic potentials. The nature of the non-Markovian process is embodied by a generalized Langevin equation under colored Gaussian noise in the form of the equation of motion
\begin{eqnarray}
m\,\dot{v}(t) &=&  -\int_{0}^{t}dt^{\prime}\,\phi(t-t^{\prime})\, v(
t^{\prime})-k_1\,x(t)
-k_2\,\left[x(t)-L(t)\right]-k_3\,x^3(t)+  {\xi}(t),\nonumber\\ &&\label{1}\\
\dot{x}(t) &=& v(t).\label{2}
\end{eqnarray} 
The colored Gaussian noise, $\xi(t)$, induces a memory kernel, $\phi(t-t^{\prime})$, consequently turning the dynamics intrinsically non-Markovian. The mechanical behavior induced by such kernel is quite unusual and instructive. We study it carefuly in appendix~\ref{appA}. Such noise is characterized by
\begin{eqnarray}
\langle \xi(t)\rangle & = &0, \\
\langle \xi(t) \xi(t^{'})\rangle & =& \frac{\gamma\,T}{\tau} e^{-\frac{|t-t^{\prime}|}{\tau}}, \\
\phi(t-t^{\prime}) &=& \frac{\langle \xi(t) \xi(t^{'})\rangle}{T}.
\end{eqnarray}
The external driving force protocol is given by~\cite{Morgado2010}
\begin{equation}
    L(t) = L_0\left(1- e^{-\frac{t}{\lambda}}\right).\label{LProtocol}
\end{equation} 

Although it can be shown that Jarzynski equality ($JE$) is valid for non-Markovian dynamics~\cite{Speck2007}, even when considering models more general than those described by generalized Langevin equations, there are many facets of this problem that deserve a careful look, and the present approach is helpful. Thus, to verify $JE$ for such dynamics, we will follow a series of steps described below:
\begin{itemize}
\item The system is initially at equilibrium with a reservoir at temperature $T$. The initial conditions ($x_0,v_0$) are consequently given by the Boltzmann-Gibbs distribution (with $L(0)=0$); 

\item At $t=0^{+}$, an external force is applied, causing a displacement described by the protocol $L(t) = L_0\,\left(1-e^{-t/\lambda}\right)$ and realizing external work $W$ over the system up to  $t=\tau$ (both the form of $L(t)$ and the finite value of $\tau$ were chosen without loss of generality and can be generalized to more complicated forms of $L(t)$ and $\tau\rightarrow\infty$);

\item The process above is repeated many times, measuring the work $W$ each run, and the non-equilibrium average $\left<e^{-\beta\,W}\right>$ is computed;

\item The equilibrium free-energies are computed for cases $F(0)\equiv F(L(0))$ and $F(\tau) \equiv F(L(\tau))$, yielding  $\Delta F = F(\tau)-F(0)$.

\end{itemize}
Consequently, $JE$ reads
\begin{equation}
 \left<\exp\left\{-\beta\,W\right\}\right> =  \exp\left\{-\beta\,\Delta F\right\}.\label{JE}
\end{equation}
The JE is an important constraint that our simulation results must obey.  We are specially interested in the change of the instantaneous entropy~\cite{Esposito2011,Parrondo2015} during the process due to the non-Markovian nature of the dissipation. In the next section we exploit the exactly solvable case of $k_3=0$. Before that we look at a special case where a seemingly violation of the 2$^{nd}$ law of Thermodynamics occur. 

\subsection{A Free-lunch}
 We can construct a simple case of a rare event (a.k.a. a free-lunch - FL) of the Jarzynski protocol, similar to the one in reference~\cite{JarzynskiPRE2006}, by studying the expansion of a system formed by an ideal gas ($N$ particles) bound by thermally conducting walls, always in contact with a thermal reservoir at temperature $T = \beta^{-1}$, and a thermally insulated piston, controlled by some given external force protocol. 

\begin{figure}
    \centering
     \includegraphics[height=5.0in,width=7.5in,angle=0]{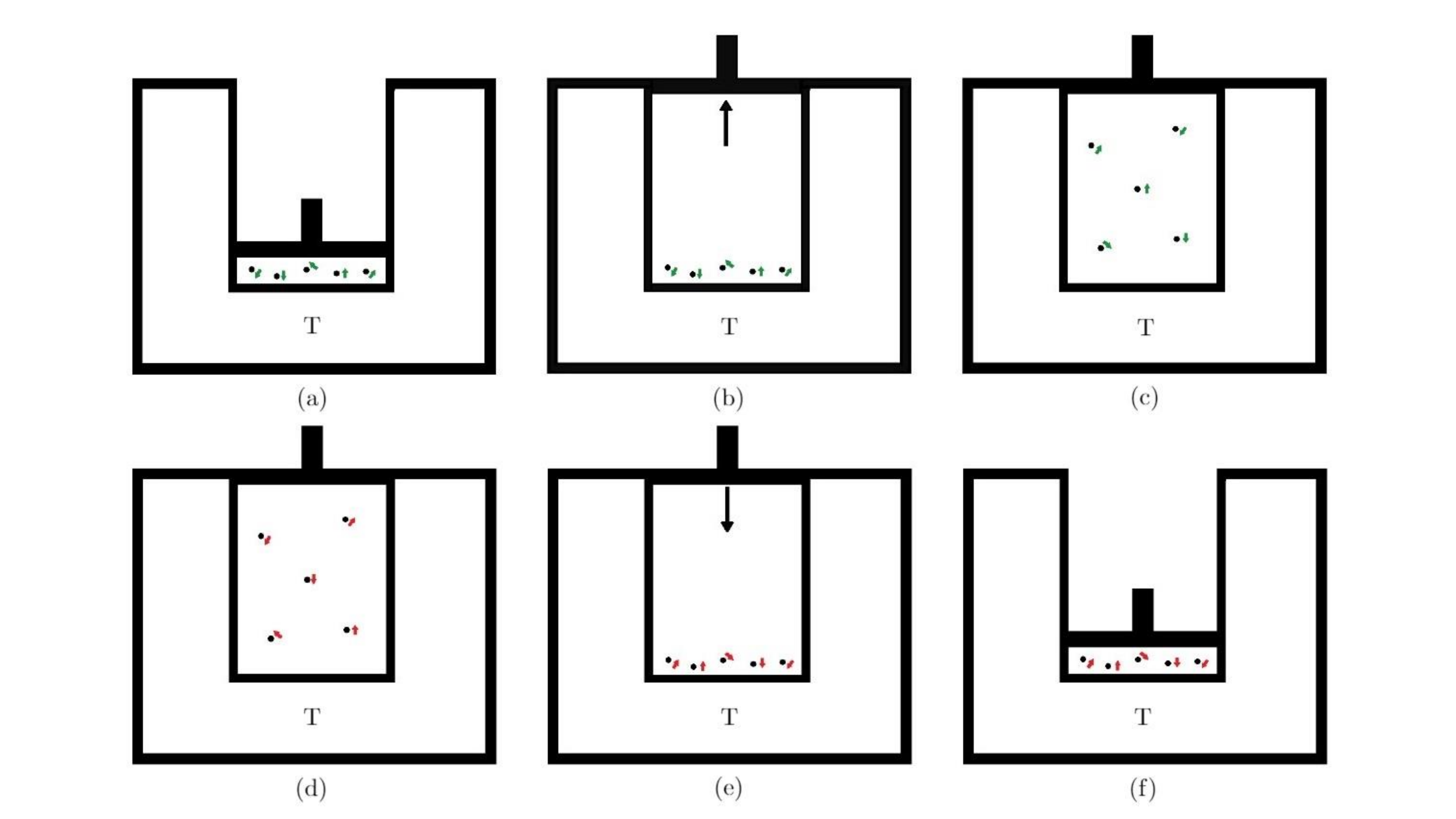}
    \caption{In (a), (b) and (c), we work out the steps that will be useful for constructing the protocol for an ideal gas example of a rare event (a free lunch). The velocities of the particles are schematically represented by green arrows in these pictures. In (a), at $t=0$, the N-particle ideal gas is at an equilibrium microstate $\mu_0$ at temperature $T$ and volume $V_0$.  We then move the piston very fast, faster than any particle of the gas, in such a way that at time $t=\tau_1 \ll 1$ the piston has reached a volume of $100 V_0$, which is depicted in (b). We then let the system thermalize for a very long time $\tau_2 \gg 1$, where it reaches a microstate $\mu_f$, an equilibrium configuration, at $t=\tau_1+\tau_2$, shown in (c). In (d), (e) and (f) the protocol, amd the rare state, are defined (now the velocities are represented in red). basically the piston protocol chosen for the problem corresponds to the inverse of the piston motion in (a)-(c). at time $t^{\prime}=0$, we start from a given equilibrium microstate at temperature $T$ and volume $100 V_0$, represented in (d). Since the N-particle ideal gas is going to eventually be violently compressed ($100 V_0 \rightarrow V_0$) the dissipated work shall be huge for any ``normal'' microstate. However, if we pick as our initial microstate $\overline{\mu_f}$, the final microstate of (c) (with the velocities inverted), we let it evolve up to $t^{\prime}=\tau_2$, represented in (e). We then compress the piston very rapidly so that at time $t^{\prime}=\tau_2+\tau_1$ the piston and system reach the equilibrium configuration represented by (f).} 
    \label{freelunch}
\end{figure}

If the gas, initially (at $t=0$) is considered to be in an equilibrium macro-state $M_0$ at volume $V_0$ and temperature $T$, we can chose a particular micro-state $\mu_0 = \left\{{\bf r}^N_0,{\bf p}^N_0\right\}\in M_0$ as our initial condition. Next, the piston is moved fast, at velocity $v_{piston}$ (faster than any of the gas particles, hence no gas-piston contact during the present expansion), from its initial position to a final position, at time $t = \tau_1$, such that $V(\tau_1) = V = 100 V_0$. It should be clear that no work is done by the piston upon the gas in such case. After a long time interval such that $\tau_2 \gg 1 \gg \tau_1$, the system thermalizes at volume $V$. Thus, when $t = \tau_1 + \tau_2$, the system is at a micro-state $\mu_f=\left\{{\bf r}^N_{t=\tau_1 + \tau_2},{\bf p}^N_{t=\tau_1 + \tau_2} \right\}  =\left\{{\bf r}^{*N},{\bf p}^{*N} \right\}  \in M_f$, where $M_f$ is an equilibrium macro-state at at volume $V$ and temperature $T$. We now define the two phase protocol to be used for the $JE$ as:
\begin{itemize}
    \item Start with the gas at thermal equilibrium ($100 V_0,T$), at the macro-state $M_f$ described earlier;
    \item Phase 1: from $t=0$ up to $t=\tau_2$, the piston is kept at its initial position;
    \item Phase 2: from $t=\tau_2$ up to $t = \tau_2+\tau_1$, the piston is moved very fast with the opposite velocity as above ($-v_{piston}$), such that the final volume is now $V_0$.
\end{itemize}
Obviously, the system does not receive any work during phase 1 but, in general, an enormous amount of work shall be done upon it during phase 2, in order to compress the system from $100 V_0 \rightarrow V_0.$ Thus, the typical dissipated work $W_{d}= W_{ext} - (F(V_0)-F(100 V_0))$ would be quite large.

However, let us chose the initial microscopic state of the gas, at the beginning of the protocol,  as the thermally equilibrated micro-state $\mu_i^{\prime}=\overline{\mu_f}\equiv \left\{{\bf r}^{*N},-{\bf p}^{*N} \right\}$, the micro-state obtained from the final state above by inverting all the molecular velocities, which is in the same macro-state $M_f$ (two micro-states, with all their particle momenta inverted, are both equilibrium micro-states whenever one of them is, due to the detailed balance property~\cite{livro_degroot}). Consequently, the gas is going to clearly reverse its trajectory and spontaneously evolve towards the final state $\mu_f^{\prime}= \overline{\mu_0}\equiv \left\{{\bf r}^N_0,-{\bf p}^N_0\right\}$. During phase 1 of the protocol, the external work shall be null, and during phase 2 the gas will have no contact with the piston, as both are retracing back their trajectories of the expansion. So, $W_{ext}=0$ in phase 2, given that the initial equilibrium state $\mu_i^{\prime}=\overline{\mu_f}$ has been chosen. The final state, $\mu_f^{\prime}=\overline{\mu_0}$ is a thermally equilibrated system at $(T,V_0)$. In this case, the dissipated work can be easily shown to be  
\[
W_{d}^{(FL)} = 0 -  \Delta F =  T \Delta S = -  N T \ln\left(\frac{V}{V_0} \right) = -  N T \ln 100,
\]
yielding
\[
e^{-\beta W_{d}^{(FL)}} = e^{N \ln 100} = 100^{N}\gg 1,
\]
which is a very large contribution to the average of $e^{-\beta W}$, in the $JE$. Micro-states such as $\mu_i^{\prime}$ are exceedingly rare, and their contribution to the instantaneous averages may be quite small, but their effect on the subsequent entropy behavior can be huge due to their effect of spontaneous entropy decreasing  which is a very large contribution to the average of $e^{-\beta W}$ in the $JE$,  and ultimately guarantees the validity of the 2nd law of Thermodynamics in its Free Energy version. Micro-states such as $\mu_i^{\prime}$ are exceedingly rare, and their contribution to the instantaneous averages may be quite small, but their effect on the subsequent entropy behavior can be huge due to their effect of spontaneous entropy decreasing. 

The mechanical properties of the non-Markovian dissipation are important enough to the understanding of the entropy behavior. In Appendix~\ref{appA} we study the damping of the initial velocity of the Brownian particle for the low and high dissipation regimes.  


\section{Non-Markovian linear model}
\subsection{Energy}

The dynamics of the system is given by the
generalized Langevin equation defined before, when $k_3=0$, where the potential energy for the system is given by 
\begin{equation}
    U(x) = \frac{k_1}{2}\,x^{2} + \frac{k_2}{2}\,\left[x-L(t)\right]^{2}.\label{harmpotenerg}
\end{equation} 
Consequently, the total energy of the system is given by the Hamiltonian 
\begin{equation}
H = {\frac {m{v}^{2}}{2}} + \frac{k_1\,x^2 }{2}  + \frac{k_2 }{2}  \left( x- L(t)\right) ^{2}.\label{kinharmpotenerg}
\end{equation} 

The process starts with  the system at equilibrium with a thermal reservoir at temperature $T$, where the initial conditions ($x_0,v_0$) are  Boltzmann-Gibbs distributed. Then, at $t=0$, an external force is applied causing the  displacement of the piston given by $L(t) = L_0\,\left(1-e^{-t/\lambda}\right)$, and also doing work on the system, up to the instant the protocol stops, $t=\tau$. Both the form of $L(t)$ and the finite value of $\tau$ where chosen without loss of generality. It is important to notice that the time-scale $\lambda$ can be set to any positive value, with $\lambda\rightarrow\infty$ corresponding to a reversible thermodynamic process.

\subsection{Probability distribution and Free-energy}
In order to obtain the non-equilibrium steady-state (NESS) probability distribution , we can start the system at any initial condition $(x_0, v_0)$ at $t_0$, and then obtain the instantaneous distribution at time $t_f$~\cite{Soares-Pinto2006,Morgado2010}. Taking $t_f\rightarrow\infty$ it is possible to obtain the NESS probability distribution. Despite the fact that the dissipation term induces non-Markovianity, the stationary  probability distribution $p_S(x,v,L)$, given the piston position $L$, is still in the Boltzmann-Gibbs form
\begin{eqnarray}
p_{S}(x,v,L) &=&\frac{\sqrt{(k_1+k_2)\,m}}{
2\,{\pi }\,{T}}\,{\rm e}^{-{\frac {m{v}^{2}}{2T}}- \frac{\left( k_1+k_2 \right)}{2T}  \left( x-{\frac {k_2 L}{k_1+k_2}} \right) ^{2}}.\label{pdfeqlinear}
\end{eqnarray}
Consequently, the initial equilibrium distribution $p_{0}(x_0,v_0) = p_{S}(x_0,v_0,L=0)$ reads
\begin{eqnarray}
p_{0}(x_0,v_0)  &=&  p_S(x_0,v_0,L=0) \nonumber\\
&=&\frac{\sqrt {(k_1+k_2)\,m}}{
2\,{\pi }\,{T}}\,{\rm e}^{-{\frac {m{v}_{0}^{2}}{2T}}- \frac{\left( k_1+k_2 \right)\,x_{0}^{2}}{2T} }.\label{pdfeqlinear0}
\end{eqnarray}

From the stationary distribution $p_S(x,v,L)$ we can obtain the equilibrium Helmholtz free energy $F(T,L)$ directly since the equilibrium entropy is given by
\[
S(T,L) = - \int \,dx\,dv\, p_S(x,v,L) \ln p_S(x,v,L), 
\]
and the internal energy is 
\[
U(T,L) = \int \,dx\,dv\, p_S(x,v,L) \, H(x,v).
\]
Thus, combining these results to find the free energy ($F = U - T\,S$) we obtain
\begin{displaymath}
F(T,L)= \frac{k_1\,k_2}{k_1+k_2}\, \frac{L^2}{2} +T\,\ln\left(\frac{\sqrt{(k_1+k_2)\,m}}{2\,{\pi }\,{T}}\right),
\end{displaymath}
which is the same as in Ref.~\cite{Morgado2010}, where it was derived for the Markovian harmonic model. Then, the free-energy difference is just 
\begin{equation}
    \Delta F = F(L_f,T) - F(0,T) = \frac{k_1\,k_2}{k_1+k_2}\, \frac{L_f^2}{2} = \left(\frac{1}{k_1}+\frac{1}{k_2}\right)^{-1}\, \frac{L_f^2}{2}.
\end{equation}

\subsection{Generating Function}
We need to construct the generating function $\left<\overline{\exp\left\{-i\,u\, W_{\theta}\right\}\} }\right>$ for the work function $W_{\theta}$, the external work done upon the system from $t=0$ to $t=\theta$, where $\overline{F}$ is the average over the initial conditions and $\left< {F}\right>$ is the average over the noise, both for any given function $F$. The cumulant generating function (CGF) is then $\ln \left<\overline{\exp\left\{-i\,u\, W_{\theta}\right\}\} }\right>$. Hence, the Jarzynski equality can be obtained through the analytic continuation, i.e., $u = -\frac{i}{T}$, of the CGF
\begin{equation}
G(u)\equiv \ln\left< \overline{\exp\left\{-i\,u\, W_{\theta}\right\}\} }\right>=
\sum_{n=1}^{\infty}\frac{(-i\,u)^n}{n!}
\left<\overline{W_{\theta}^n}\right>_c.\label{GudeWext}
\end{equation}
Due to the linearity of the model, the distribution is forcibly Gaussian and an exact solution ensues.  Only the average and the variance of $W_{\theta}$ will be non-zero. Observe that the Jarzynski equality should occur at $u = -\frac{i}{T}$, since 
\[
\exp\left\{G\left(-\frac{i}{T}\right)\right\} = \left<\overline{\exp\left\{-\frac{W_{\theta}}{T} \right\} }\right>,
\]
which, due to the $JE$, we must have 
$G\left(-\frac{i}{T}\right) \equiv -\frac{\Delta F}{T}$.

\subsection{Work cumulants}
The work expression in Eq.~(\ref{GudeWext}) corresponds to the external work done upon the system, given by the $L(t)$ protocol. It reads
\begin{eqnarray}
W_{\theta} &=&-k_2\int_{0}^{\theta}dt \,
\frac{dL}{dt}\left(x(t)-L(t)\right)\nonumber\\
&=& \Delta\,U-k_2\int_{0}^{\theta}dt \,
\frac{\partial\,L(t)}{\partial\,t}x(t),\nonumber\\
&\equiv & \Delta\,U+ I_{\theta},\label{work}
\end{eqnarray}
with $\Delta\,U=k_2L_0^2/2$. It is the
coupling of $x(t)$ and $\frac{\partial\,L(t)}{\partial\,t}$ that will
give rise to the irreversible work loss. Thus, lets rewrite the integrals in the form  
\begin{eqnarray*}
I_{\theta} &=& - k_2\int_{0}^{\theta}dt \,
\frac{\partial\,L(t)}{\partial\,t}\,x(t) \\
&=&   - k_2\int_{0}^{\theta}dt \,\frac{\partial\,L(t)}{\partial\,t}\int_{0}^{\infty}\,dt_1\,\delta(t-t_1)  \,x(t_1) \\
&=&   - \frac{k_2\,L_0}{\lambda}\int_{0}^{\theta}dt \,e^{-\frac{t}{\lambda}}\int_{0}^{\infty}\,dt_1\,   \int_{-\infty}^{\infty}\frac{dq_1}{2\pi}\,e^{(iq_1+\epsilon)(t-t_1)}\,x(t_1)  \\
&=&    \frac{k_2\,L_0}{\lambda}\,   \int_{-\infty}^{\infty}\frac{dq_1}{2\pi}\, \frac{\left(e^{-\left[\frac{1}{\lambda}-(iq_1+\epsilon)\right]\theta} - 1\right) }{\frac{1}{\lambda} -(iq_1+\epsilon)}\,\tilde{x}(iq_1+\epsilon),
\end{eqnarray*}
where $\tilde{x}(iq_1+\epsilon)$ corresponds to the Laplace-Fourier Transform of the position.

The cumulants of $W_{\theta}$ will be given by
\begin{eqnarray}
 \left<W_{\theta}\right>_c &= &  \Delta U + \left<I_{\theta}\right>_c, \label{Wc1} \\
 \left<W_{\theta}^2\right>_c &= &   \left<I_{\theta}^2\right>_c, \label{Wc2} \\
 \left<W_{\theta}^{n\geq 3}\right>_c &= & 0. \label{Wcmaiorque2}
\end{eqnarray}
Consequently, the CGF in Eq.~(\ref{GudeWext}) can be easily derived as
\begin{equation}
G(u)\equiv \ln \left< \overline{\exp\left\{-i\,u\, W_{\theta}\right\}\} }\right>= -iu\,\Delta U\,+ \ln \left< \overline{\exp\left\{-i\,u\, I_{\theta}\right\}\} }\right>= -iu\,\Delta U -i\,u\,
\left<\overline{I_{\theta}}\right>_c -
\frac{u^2}{2}\,
\left<\overline{I_{\theta}^2}\right>_c.
\end{equation}
The calculations, for both first and second order cumulants, are rather cumbersome but straightforward. These cumulants are obtained exactly for the linear harmonic model. In the following, three important quantities are $\kappa_1$ (real), and $\kappa_2=\kappa_3^*$ (complex). They are the zeroes of the factor $R(s)$, for the harmonic case, defined as
\[
\tilde{x}(s) = \frac{ \tilde{\xi}(s)}{R(s)}.
\]
We also write $\kappa_2=\kappa_R + i\,\kappa_I.$ 
The actual expressions are very long and cumbersome, but can be worked out directly without much problem. We study these expressions and their consequence in appendix~\ref{appendix-discriminant}. An interesting phase diagram arises corresponding to $\kappa_I$ being real or complex (in such case all the $\kappa_{1,2,3}$  are real). This will have important consequences for the time behavior  of the information entropy of the system. 

Due to the Gaussian property of the initial conditions, and of the time evolution equations and noise, the instantaneous distributions will be Gaussian and only the first two cumulants of the work need to be taken into account. In appendix~\ref{appendix-As} we describe the details of the calculations for the two cumulants of the work function $I_{\theta}$ below.

\subsubsection{First-order cumulant $\left<\overline{I_{\theta}}\right>_c$}
In order to calculate the cumulant of a dynamical function, say ${F}$, we first take the noise average $\left<{F}\right>$, then we take the average over the initial conditions $\overline{{F}}$. The only first cumulant contribution is 
\begin{eqnarray*}
\left<\overline{I_{\theta}}\right>_c &=& \overline{\left<\left( \frac{k_2\,L_0}{\lambda}\,   \int_{-\infty}^{\infty}\frac{dq_1}{2\pi}\, \frac{\tilde{x}(iq_1+\epsilon)\,\left(e^{-\left[\frac{1}{\lambda}-(iq_1+\epsilon)\right]\theta} - 1\right) }{\frac{1}{\lambda} -(iq_1+\epsilon)} \right)\right>_c} \\
&=&
\frac{1}{2}\,{k_{{2}}}^{2}{L_{{0}}}^{2} \left( {{\rm e}^{-{\frac {2\,\theta}{
\lambda}}}}-1 \right)  \frac{\left( -\tau+\lambda \right)}{ {\lambda}^{2}\,{m}\,{\tau} \left( 1+\kappa_{{1}}\lambda \right) \left( 1+
\kappa_{{2}}\lambda \right) \left( 1+\kappa_{{3}}\lambda
 \right)} \\
 &-&  {k_{{2}}}^{2}{L_{{0}}}^{2} \left( {{\rm e}^{-{\frac {
\theta}{\lambda}}}}-1 \right) \frac{1}{m\,\tau\,\kappa_{{1}}\,\kappa_{{2}}\,\kappa_{{3}}}\\
 &-& {k_{{2}}}^{2}{L_{{0}}}^{2}
 \left( {{\rm e}^{{\frac { \left( \kappa_{{1}}\lambda-1 \right) \theta
}{\lambda}}}}-1 \right)  \frac{\left( 1+\tau\,\kappa_{{1}} \right)}{m\,\tau\,\kappa_{{1}} \left( \kappa_{{1}}-\kappa_{{2}}
 \right) \left( \kappa_{{1}}-\kappa_{{3}} \right) \left( {
\kappa_{{1}}}^{2}{\lambda}^{2}-1 \right)}\\
&+&  {k_{{2}}}^{2}{L_{{0}}}^
{2} \left( {{\rm e}^{{\frac { \left( \kappa_{{2}}\lambda-1 \right) 
\theta}{\lambda}}}}-1 \right)  \frac{\left( 1+\tau\,\kappa_{{2}} \right)}{ m\,\tau\,\kappa_{{2}} \left( \kappa_{{1}}-\kappa_{{2}}
 \right) \left( \kappa_{{2}}-\kappa_{{3}} \right) \left( {
\kappa_{{2}}}^{2}{\lambda}^{2}-1 \right)}\\
&-& {k_{{2}}}^{2}{L_{{0}}}^
{2} \left( {{\rm e}^{{\frac { \left( \kappa_{{3}}\lambda-1 \right) 
\theta}{\lambda}}}}-1 \right)  \frac{\left( 1+\tau\,\kappa_{{3}} \right)}{m\,\tau\,\kappa_{{3}} \left( \kappa_{{1}}-\kappa_{{3}}
 \right) \left( \kappa_{{2}}-\kappa_{{3}} \right) ^{-1} \left( {
\kappa_{{3}}}^{2}{\lambda}^{2}-1 \right)}
\end{eqnarray*}

\subsubsection{Second-order cumulant $\left<\overline{I_{\theta}^2}\right>_c$}
Like the first order cumulant, we  break the calculation into the following parts:
\begin{eqnarray*}
\left<\overline{I_{\theta}^2}\right>_c & =& \mathcal{A}_1+\mathcal{A}_2+\mathcal{A}_3+\mathcal{A}_4+\mathcal{A}_5, 
\end{eqnarray*}
where the expressions for the $\mathcal{A}_i$ terms can be found in Appendix \ref{appendix-As}.

The verification of $JE$ comes from the calculation
\[
G\left(u=-\frac{i}{T}\right)= \ln \left< \overline{\exp\left\{- \frac{W_{\theta}}{T}\right\} }\right>= -\,\frac{\Delta U}{T} -\frac{\left<\overline{I_{\theta}}\right>_c}{T}+\frac{\mathcal{A}_1+\mathcal{A}_2+\mathcal{A}_3+\mathcal{A}_4+\mathcal{A}_5}{2T^2}.
\]
The lengthy expressions above can be easily simplified yielding
\[
\Rightarrow G\left(u=-\frac{i}{T}\right) =  \frac{k_1\,k_2}{k_1+k_2}\, \frac{L_{\theta}^2}{2} = -\frac{\Delta F}{T},
\]
showing that the Jarzynski equality holds exactly for the non-Markovian linear case.

\section{Entropy and the Jarzynski Equality: analytical results}
We shall exploit the properties of the entropy change of the linear model during the action of the protocol. The Gaussian property, coupled with the linearity of the model reveals some surprising consequences as those shown in the following.

\subsection{Entropy budget}
In the context of applying information theory to the analysis of non-equilibrium systems, the use of the so called out-of-equilibrium notions for entropy and free-energies becomes quite useful~\cite{Esposito2011,Parrondo2015}. For instance, it allows to calculate the maximum work extractable from a system undergoing a non-equilibrium process. Under this perspective, in the present case, lets start assuming that the system is always in contact with a thermal bath at constant temperature $T$. Being the non-equilibrium \underline{informational entropy} defined as ($k_B=1$)
\begin{equation}
    S_{\rm sys}(t) = -\int\,dx\,dv\,\rho(x,v,t) \ln \rho(x,v,t),\label{Sdet}
\end{equation}
where the internal energy is 
\begin{equation}
    U(t) = \int\,dx\,dv\,\rho(x,v,t) \, H(x,v),\label{Edet}
\end{equation}
and the out-of-equilibrium  free-energy reads
\begin{equation}
    F(t) = U(t)-T\,S(t).\label{Fdet}
\end{equation}

For the reservoir, the corresponding change of entropy is given by by the negative of the amount of heat flowing towards the system, $j_Q$, since the equilibrium reservoir does not produce entropy. We have
\begin{equation}
    \Delta S_R(t) = - \frac{1}{T}\int_{0}^{t}ds\,j_Q(s).
\end{equation}
The heat flow expression can be easily obtained as~\cite{Nascimento2020} 
\begin{equation}
    j_Q(s) = \xi(s)\,v(s)  -\int_{0}^{s}dt^{\prime}\,\phi(s-t^{\prime})\, v(t^{\prime})\, v(s),
\end{equation}
yielding
\begin{equation}
     \Delta S_R =  \frac{1}{T}\int_{0}^{t}ds\,\left(\int_{0}^{s}dt^{\prime}\,\phi(s-t^{\prime})\, v(t^{\prime})\, v(s)-\xi(s)\,v(s) \right).\label{reservoirentropychange}
\end{equation}
The change in the system entropy can be found from
\begin{equation}
    \Delta S_{\rm sys}(t) = -\int\,dx\,dv\,\left(\rho(x,v,t) \ln \rho(x,v,t)-\rho(x,v,0) \ln \rho(x,v,0)\right).\label{deltaS}
\end{equation}
Consequently, the total entropy change, i.e., for system and thermal reservoir, is the sum of the therms in Eqs.~(\ref{reservoirentropychange}) and~(\ref{deltaS}):
\begin{equation}
    \Delta S_{tot} (t) = \Delta S_{\rm sys}(t)+ \Delta S_R(t).
\end{equation}
Next we are going to obtain the exact results for the linear case scenario.

\subsection{Entropy calculation: the harmonic case}
\label{section-entropy_calculation}
At the beginning of the protocol ($t=0$), the system is in thermal equilibrium at temperature $T$ with the reservoir and, thus, the initial probability distribution $p_0(x_0,v_0)\equiv p(x_0,v_0,t=0)$ is Gaussian distributed, due to the harmonic nature of the elastic interactions and of the quadratic form for the kinetic energy. 
%
%
%
Hence
\begin{eqnarray*}
p_0(x_0,v_0)  &=& \frac{\sqrt {(k_1+k_2)\,m}}{
2\,{\pi }\,{T}}\,{\rm e}^{-{\frac {m{v}_{0}^{2}}{2T}}- \frac{\left( k_1+k_2 \right)\,x_{0}^{2}}{2T} }.
\end{eqnarray*}

There is an interesting result to be discussed for the harmonically bound particle. Starting from the linear version ($k_3=0$) of Eq.\ref{1},
\begin{eqnarray}
 m\,\dot{v}(t) =  -\int_{0}^{t}dt^{\prime}\,\phi(t-t^{\prime})\, v(
t^{\prime})-k_1\,x(t)
-k_2\,\left[x(t)-L(t)\right] +  {\xi}(t). \label{1-linear}
\end{eqnarray}
In appendix~\ref{appA}, we study the mechanical consequences of the non-Markovian memory kernel, which tell us that the competition between the memory time-scale and the dissipation time-scale can lead to oscillations of velocity in time. Hence, we might expect that the statistical consequence of the mechanical oscillations would be entropy oscillations over time.

Due to the linearity of the stochastic equations of motion, we can use the Green's function approach~\cite{Morgado2010}. The particular solution (which has the noise function $\xi(t)$ and the protocol for $L(t)$ as the source terms) and the homogeneous one (which depends on the initial quantities $x_0$ and $v_0$) are combined below: 
	\begin{eqnarray}
	x(t) = \int_0^t d t' g(t-t') \Big[ \xi(t') + k_2 L(t') \Big] + x_0 f(t) + m\,v_0\,g(t), \\
	v(t) = \int_0^t d t' \dot{g}(t-t') \Big[ \xi(t') + k_2 L(t') \Big] + x_0 \dot{f}(t) + m\,v_0\,\dot{g}(t),
	\end{eqnarray}
where the Green function $g(t)$ and the auxiliary function $f(t)$ are given by
	\begin{eqnarray}
	g(t) = \lim_{\epsilon \to 0} \int_{-\infty}^\infty \frac{dq}{2 \pi} \frac{e^{(iq+\epsilon)t}}{R(i q + \epsilon)} ~~ , ~~ 	f(t) = \lim_{\epsilon \to 0} \int_{-\infty}^\infty \frac{dq}{2 \pi} \frac{m \,(iq + \epsilon) + \tilde{\phi}(i q + \epsilon)}{R(i q + \epsilon)}e^{(iq+\epsilon)t},
	\end{eqnarray} 
with  $R(s) = m s^2 + \tilde{\phi}(s) s + k_1 + k_2$, $\tilde{\phi}(s)$ being the Laplace transform of the damping kernel $\phi(t)$, and the former can be rewritten as
\begin{eqnarray}
 R(s) &=& m s^2 + \frac{\gamma s}{1 + \tau s} + k_1 + k_2 = \frac{m \tau s^3 + m s^2 + ( \gamma + \tau(k_1 + k_2) ) s + k_1 + k_2}{1 + \tau s} \nonumber \\
 &=& \frac{m (s - \kappa_1)(s - \kappa_2)(s - \kappa_3)}{1 + \tau s},
\end{eqnarray}
where the numerator is in a more compact form using its roots $\kappa_1$, $\kappa_2$ and $\kappa_3$. The values of $\kappa$ will depend on the system parameters, a more detailed analysis of the possible results can be found in Appendix \ref{appendix-discriminant}.

The next step is to solve for the cumulants of the instantaneous distribution, namely the averages and variances, of $x(t)$ and $v(t)$. Since $\langle \xi(t) \rangle = 0$, the averages for the position and the velocity of the particle can be calculate from
	\begin{eqnarray}
	\mu_x(t) &=& \big\langle x(t) \big\rangle = \int_0^t d t' g(t-t') k_2 L(t')  + \langle x_0 \rangle f(t) + \langle v_0 \rangle\,m,g(t)  \\
	\mu_v(t) &=& \big\langle v(t) \big\rangle = \int_0^t d t' \dot{g}(t-t') k_2 L(t')  + \langle x_0 \rangle \dot{f}(t) + \langle v_0 \rangle\,m\, \dot{g}(t), 
	\end{eqnarray}
where, due to the symmetrical nature of the initial conditions, $\langle x_0 \rangle = \langle v_0 \rangle=0$.
The only surviving contribution to the averages is a (deterministic) term corresponding to the deterministic integral of $L(t^{\prime}).$ That contribution is associated to the changes for the equilibrium averages of $x$ due to the moving of the so called piston, i.e., the free extremity of the spring linking the Brownian particle to the external system.

\begin{figure}
\begin{center}
    \includegraphics[width=0.45\textwidth]{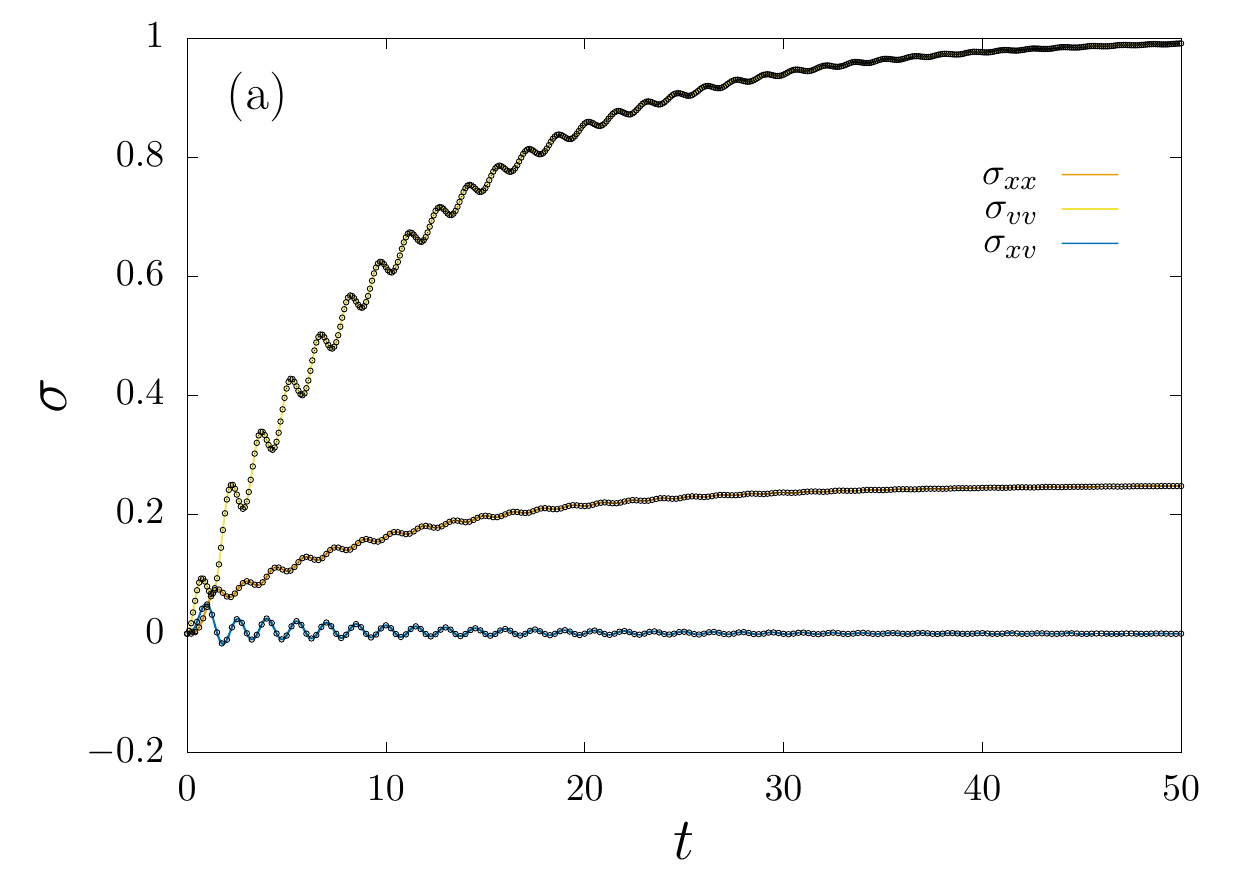}
    \includegraphics[width=0.45\textwidth]{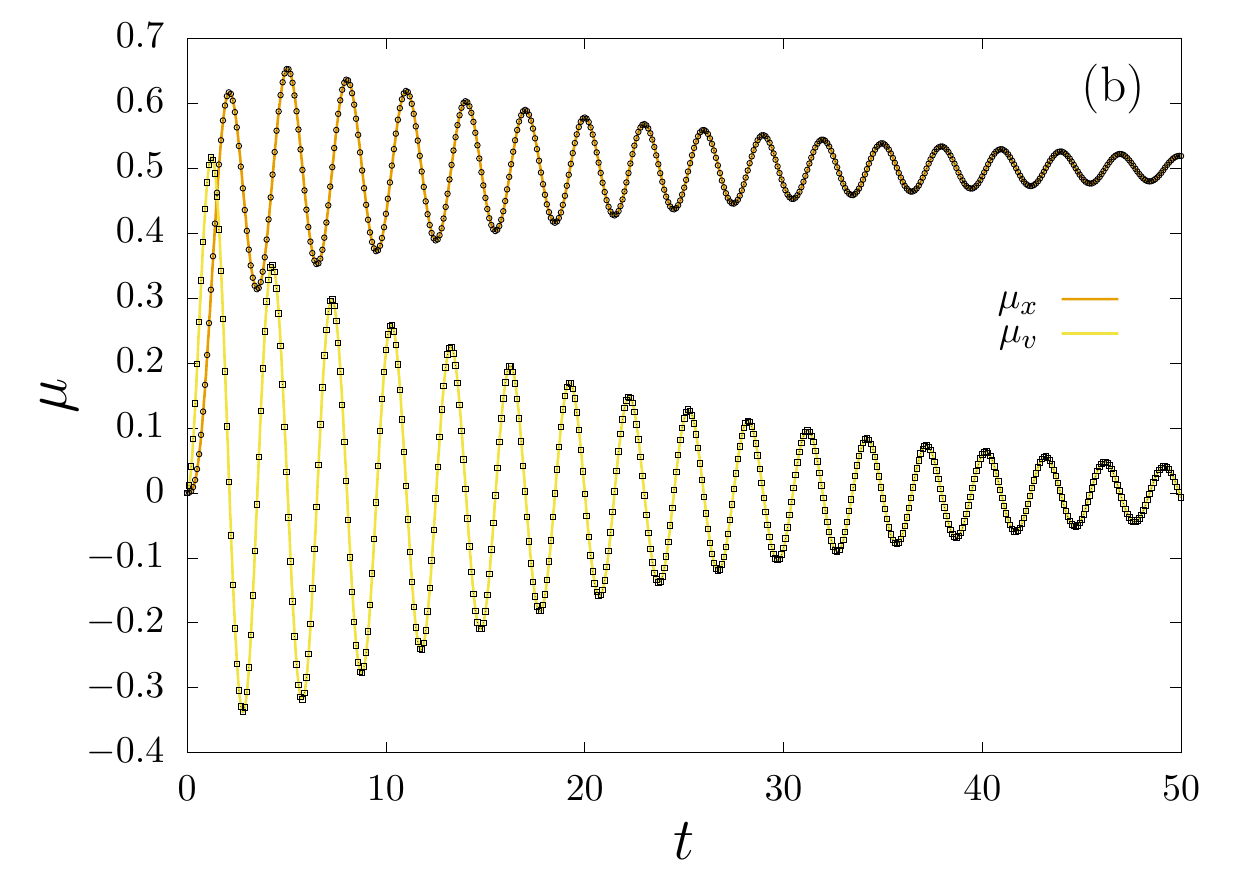}
    \caption{ We compare the numerical results (points) with the theoretical predictions (solid lines) for the $\sigma$-functions in (a) and for the $\mu$-functions in (b). For the numerical integrations we used $k_1 = k_2 = 2$, $\gamma = 0.5$ and $m = \tau = T = \lambda = L_0 = 1$, with $x_0=v_0=0$, the figure represents the process of thermalization of the system. Note that the frequency of observed oscillations in panel b) is identical for $\mu_x$ and $\mu_v$, with it's value being determined by the imaginary root $\kappa_I \simeq 2.1$. On panel a) the frequency of oscillations is a combinaton of $\kappa_I$ and $2 \kappa_I$ (with the latter being the dominant), as \ref{eq-sigmas} is proportional to square terms of $g(t)$ and $f(t)$. \label{sigma-compare}}
    \label{fig:my_label}
\end{center}
\end{figure}

Since the variances are defined by
\begin{eqnarray}
	\sigma_{xx}(t) &=& \big\langle x(t) x(t) \big\rangle - \mu_x(t)^2, \\
	\sigma_{vv}(t) &=& \big\langle v(t) v(t) \big\rangle - \mu_v(t)^2, \\
	\sigma_{xv}(t) &=& \big\langle x(t) v(t) \big\rangle - \mu_x(t) \mu_v(t),
\end{eqnarray}
it is straightforward to show that the second moments contributions due to the protocols will be identically canceled by the averages products.  This remains true regardless the protocol we use. Indeed,  after a little algebra, we can write
\begin{eqnarray}
\sigma_{xx}(t) &=& \int_0^t dt_1 dt_2 g(t-t_1) g(t-t_2) \big\langle \xi(t_1) \xi(t_2) \big\rangle + \langle x_0^2 \rangle f^2(t) + \langle v_0^2 \rangle \, m^2 g^2(t), \\
\sigma_{vv}(t) &=& \int_0^t dt_1 dt_2 \dot{g}(t-t_1) \dot{g}(t-t_2) \big\langle \xi(t_1) \xi(t_2) \big\rangle + \langle x_0^2 \rangle \dot{f}^2(t) + \langle v_0^2 \rangle \, m^2 \dot{g}^2(t),\\
\sigma_{xv}(t) &=&\int_0^t dt_1 dt_2 g(t-t_1) \dot{g}(t-t_2) \big\langle \xi(t_1) \xi(t_2) \big\rangle + \langle x_0 \, v_0 \rangle \, m \big\{ f(t)\dot{g}(t) + \dot{f}(t) g(t) \big\} + \nonumber \\
&+& \langle x_0^2 \rangle \dot{f}(t) f(t) + m^2 \langle v_0^2 \rangle \dot{g}(t)g(t), \label{eq-sigmas}
\end{eqnarray} 
where we can see that the variances  $\sigma_{xx}$, $\sigma_{xx}$, and $\sigma_{xx}$ do not depend on $L_0$ or $\lambda$. 
In Fig.~\ref{sigma-compare}, we can note the time-evolution of the variances for a system coupled to a thermal bath, given that the initial conditions are fixed at $x_0=v_0=0$. This has interesting consequences with respect to the evolution of the total entropy, as we shall see in the following. 

The instantaneous probability distribution for the system, in terms of the averages and variances above, is given by
\begin{eqnarray}
    p(x,v,t) = \frac{1}{2 \pi \sqrt{\sigma_{xx} \sigma_{vv} - \sigma_{xv}^2}} \exp \Bigg\{ - \frac{1}{2}  \frac{\sigma_{vv}(x- \mu_x)^2 + \sigma_{xx} (v - \mu_v)^2 - 2 \sigma_{xv} (x - \mu_x)(v - \mu_v)}{\sigma_{xx}\sigma_{vv} - \sigma_{xv}^2} \Bigg\}.\label{entropysystemharmonic}
\end{eqnarray}
In Eq.~\ref{entropysystemharmonic}, all the influence of the protocol is restricted to the terms $\mu_{x,y}$.

The exact form for the entropy can be obtained after the substitution of Eq.~(\ref{entropysystemharmonic}) into the Shannon form as in Eq.~(\ref{Sdet})
\begin{eqnarray}
	S(t) = \frac{1}{2} \ln \Big( \sigma_{xx}(t) \sigma_{vv} (t) - \sigma_{xv}^2(t) \Big) + \ln 2 \pi e, \label{entrpoyharmonicNM} 
\end{eqnarray}
which is completely independent of the protocol variables. In fact, for more general protocols still keeping the system harmonic and Gaussian, the elimination of the protocol related deterministic terms in the equations defining the variances related to the protocol also occur, which makes the variance independent of the protocol. Hence, the entropy obtained in Eq.~(\ref{entrpoyharmonicNM}) is independent of the protocol variables
\begin{equation}
\frac{\partial S}{\partial L_0} =\frac{\partial S}{\partial \lambda} = 0.
\end{equation}
Observe that  $\sigma_{xx}(t) \sigma_{vv} (t) - \sigma_{xv}^2(t)$ is always a positive quantity. That guarantees the entropy $S(t)$ is well defined. In the following we exploit this rather unexpected result for the non-Markovian harmonic model. 

It is important to highlight that if one starts with the system thermalized with the same temperature as that of the thermal reservoir, the variances shall keep their equilibrium values. the only effect of the protocol is to displace the averages of the Brownian variables.

\begin{figure}
\begin{center}
\includegraphics[height=3.0in,width=4.0in,angle=0]{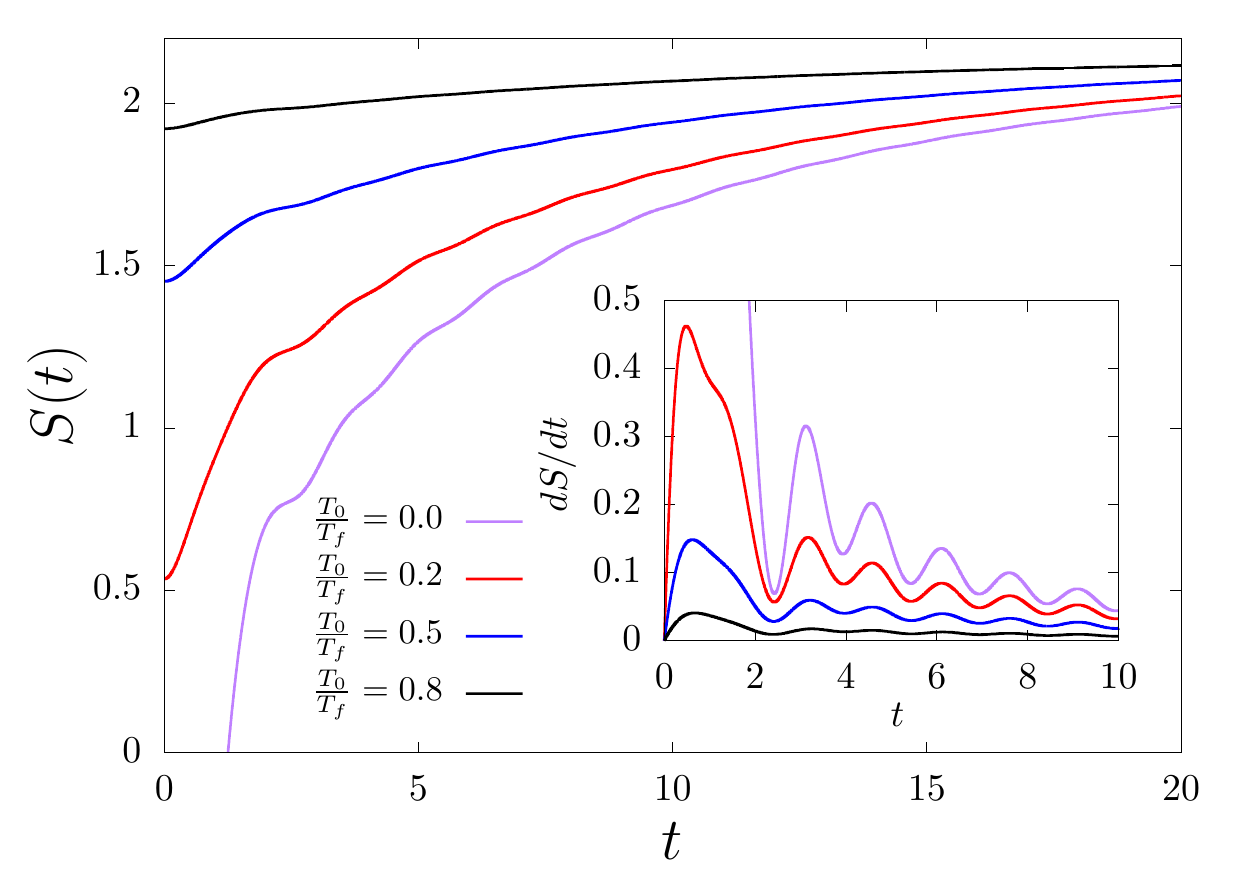}
\caption{Entropy evolution for a Brownian particle system with initial fixed conditions thermalized at various initial temperatures while the inset shows the entropy change rate. Even tough the change rate displays oscillatory behaviour, since both $g(t)$ and $f(t)$ oscillate, the entropy is still a monotonically function, increasing towards equilibrium as $dS/dt \geq 0$ for all times. The parameters used are $\gamma = 0.5$, $k_1 = k_2 = 2$ and $T_f = m= \lambda = L_0 = \tau=1$. 
}\label{entrpoyfromzerotomax}
\end{center}
\end{figure}

The non-usual entropy behavior that we observe is mostly encoded in the transient behavior of $\sigma_{xv}(t)$, which is zero at equilibrium. Correlations of velocity and position are intimately linked to the slow hydrodynamic modes, that build up in a fluid perturbed by the motion of a Brownian particle,  giving rise to the dissipative memory function underlying the present non-Markovian model. In Fig.~\ref{entrpoyfromzerotomax}, it is possible to see the evolution  towards equilibrium of the entropy for such a system. It is highly  non-trivial,  as the entropy exchange rate fluctuates very strongly (see inset of figure~\ref{entrpoyfromzerotomax}).  In all the simulations of the linear system, the set of variables used were $k_1= k_2 = 2, \gamma = 0.5, \tau=m=1.$  We stress out that in these  simulations the pulling protocol is irrelevant.

\subsection{Entropy production?}
In order for a system to produce entropy it is necessary to take that system to a non-equilibrium state, i.e., to realize non-quasi-static processes on it. Take the present pulling protocol model, unless the pulling rate is quasi-static ($\lambda\rightarrow0$), that should be enough to take it to an out-of-equilibrium state. 

However, this is not what takes place here. For the non-Markovian harmonic model with Gibbs equilibrium initial conditions, since $L(t=0)=0$, the system stays in that equilibrium state and no entropy is produced. A fast pulling protocol might be a  necessary condition for the production of entropy. But is it a sufficient one?

Let us start by clarifying the locus of entropy production in the present model. The triad system of interest (the Brownian particle and springs), external system, and thermal reservoir (see Sekimoto's book~\cite{Sekimoto2010}) characterizes well the work and heat exchanged by the system. We assume that the external system is a pure work reservoir, injecting no entropy into the system of interest. We assume it does work into the system of interest in an ordered way, with no increase of its entropy. The thermal reservoir may well exchange entropy with the system of interest, due to the exchanged heat, but it will not produce entropy itself, since we assume it to be in a state of equilibrium itself.  Indeed, this is in sharp contrast with athermal reservoirs, such as Poisson reservoirs~\cite{Kanazawa2013,Kanazawa2015}, which continuously produce entropy to preserve its non-equilibrium athermal state Hence, the only possible source for any produced entropy lies with the system of interest itself.

Consequently, the total entropy of the system and the thermal reservoir together shall vary by the amount produced in the system itself. The total entropy budget shall follow 
\begin{equation}
    \Delta S_{total}(t) = \Delta S(t) + \Delta S_R(t) = \Pi_S(s),
\end{equation}
where $S_{total}$ corresponds to the total entropy variation of system, external system and reservoir, and $\Pi_S(t)$ is the total entropy produced in the system during the interval $(0,t)$.

Since only the pulling protocol $L(t)$ would be capable of taking the system away from equilibrium~\cite{Morgado2010}, no entropy will be generated in the system as time goes on. From  Eq.~(\ref{entrpoyharmonicNM}) we have
\[\frac{\partial S}{\partial L_0} =\frac{\partial S}{\partial \lambda} =0,\]
we deduce that the result is the same if we take $\lambda\rightarrow\infty$, which is the quasi-static protocol. Hence, the system is always at equilibrium, without any entropy production, yielding
\begin{equation}
    \Pi_S(t \geq 0) = 0. \label{entropyharmonnonmarkovequaltozero}
\end{equation}

In fact, in order for the entropy production to be non-zero during the protocol, the presence of a non-harmonic potential is essential. In the following we shall study the case of a (small) non-linear term in the potential.

\subsection{Entropy oscillations and entropy backflow: linear case}
The essential effect of the presence of a memory kernel is to feed the present with information from the past, hence the non-Markovian property of the model. A possible outcome of a non-Markovian model might be information backflow, which is defined as $dS/dt<0$.
We shall see that it will be the case for the present model. However, the non-Markovian model not always leads to information backflow. 

Another interesting effect shall be called ``entropy oscillations", where we can identify the traces of a typical oscillatory behavior for the time-evolution of the entropy $S(t)$. For the present model, oscillations are present for the regimes of high-and low-dissipation, being absent for an intermediate range of the dissipation intensity.

\begin{figure}
\begin{center}
\includegraphics[height=3.0in,width=4.0in,angle=0]{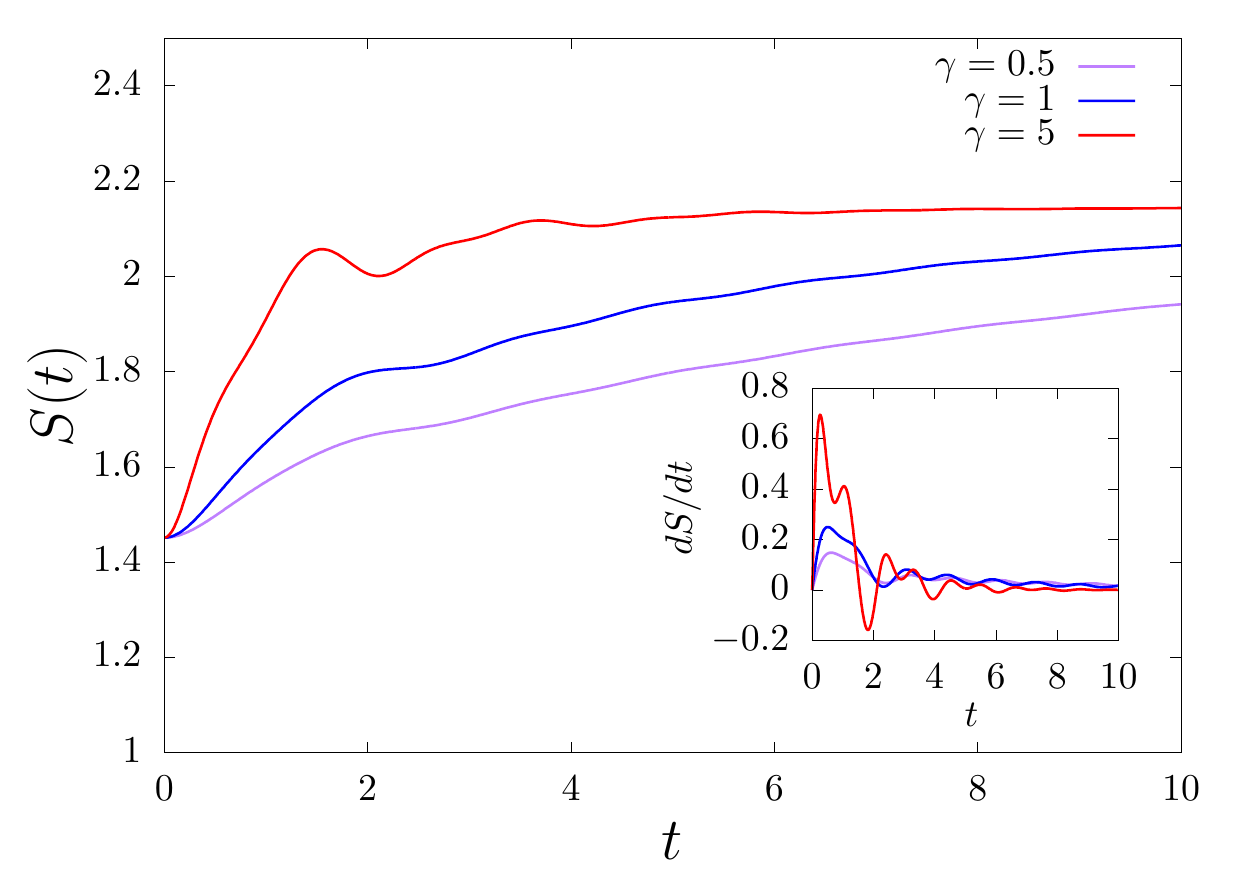}
\caption{We show the evolution of entropy for different values of the damping constant $\gamma$, while the inset shows the entropy change rate. Unlike in Fig. \ref{entrpoyfromzerotomax}, the values of $\gamma$ are high enough so that the system displays information backflow, the change rate now becomes negative and the entropy is no longer a monotonically increasing function. The parameters are the initial temperature $T_0=0.5$ and the final temperature $T_f = 1.0$, $k_1 = k_2 = 2$ and $m=\tau= \lambda = L_0 =1$. }
\label{entropyoscillationsscenarios}
\end{center}
\end{figure}

We have analyzed a few scenarios, shown in Fig.~\ref{entropyoscillationsscenarios}, where the system starts thermalized at a temperature $T_0$, which can be chosen arbitrarily. In Fig.~\ref{entropyoscillationsscenarios}, we observe that, for high enough values for the coefficient of dissipation $\gamma$, the entropy change rate becomes negative, showing oscillations for a range of time much longer than the memory time-scale $\tau = 1.$

Interestingly, the memory function acts as an information pump, recovering partially some of the information lost to the reservoir due to the information backflow ($dS/dt<0$). The information backflow effect diminishes gradually as the system reaches equilibrium, as can be seen clearly in the inset of Fig.~\ref{entropyoscillationsscenarios}.

\begin{figure}
\begin{center}
\includegraphics[height=3.0in,width=4.0in,angle=0]{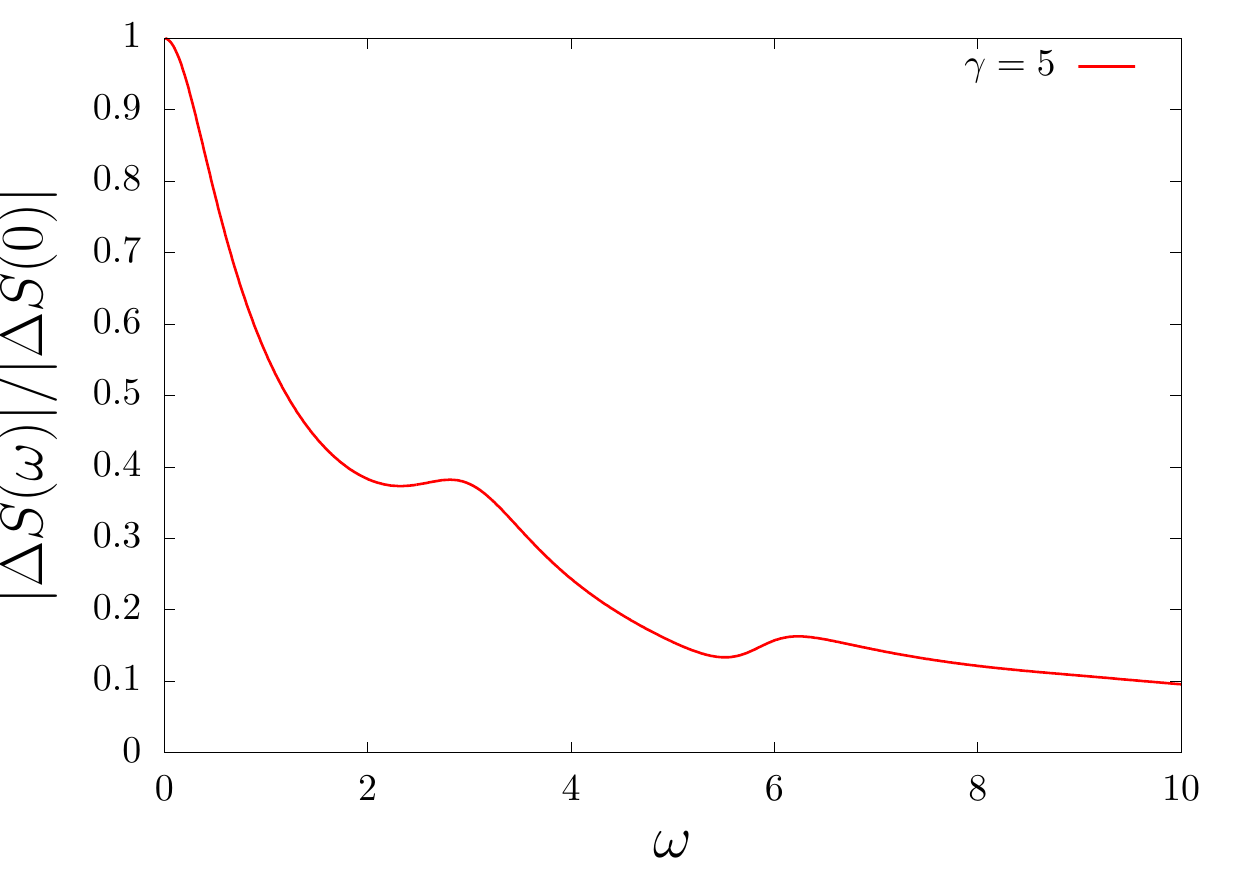}
\caption{The spectrum for $\Delta S = S(t) - S(\infty)$ of figure~\ref{entropyoscillationsscenarios} for $\gamma=5$. The large value at $\omega=0$ occurs because the entropy approaches the limit from beneath, incurring a large area. Subsequent peaks are related to multiples of the natural frequency driving the Green function $g(t)$, which for the parameters is $\kappa_I \approx 2.945$.}
\label{entropyspectrumlinear}
\end{center}
\end{figure}

The spectrum of the entropy variation of Fig.~\ref{entropyoscillationsscenarios} is plotted in Fig. \ref{entropyspectrumlinear}. There are visible peaks, belonging to the harmonic series generated for $\kappa_I =2.945$, at approximately 3 and 6. The actual value for the peaks are not exact multiples of $\kappa_i.$ The spectral analysis of the entropy as a function of time ferrets out the oscillating behavior of the variances quite clearly. In appendix~\ref{appendix-discriminant} we study the behavior of the $\kappa$'s.

\begin{figure}
\begin{center}
\includegraphics[height=3.0in,width=4.0in,angle=0]{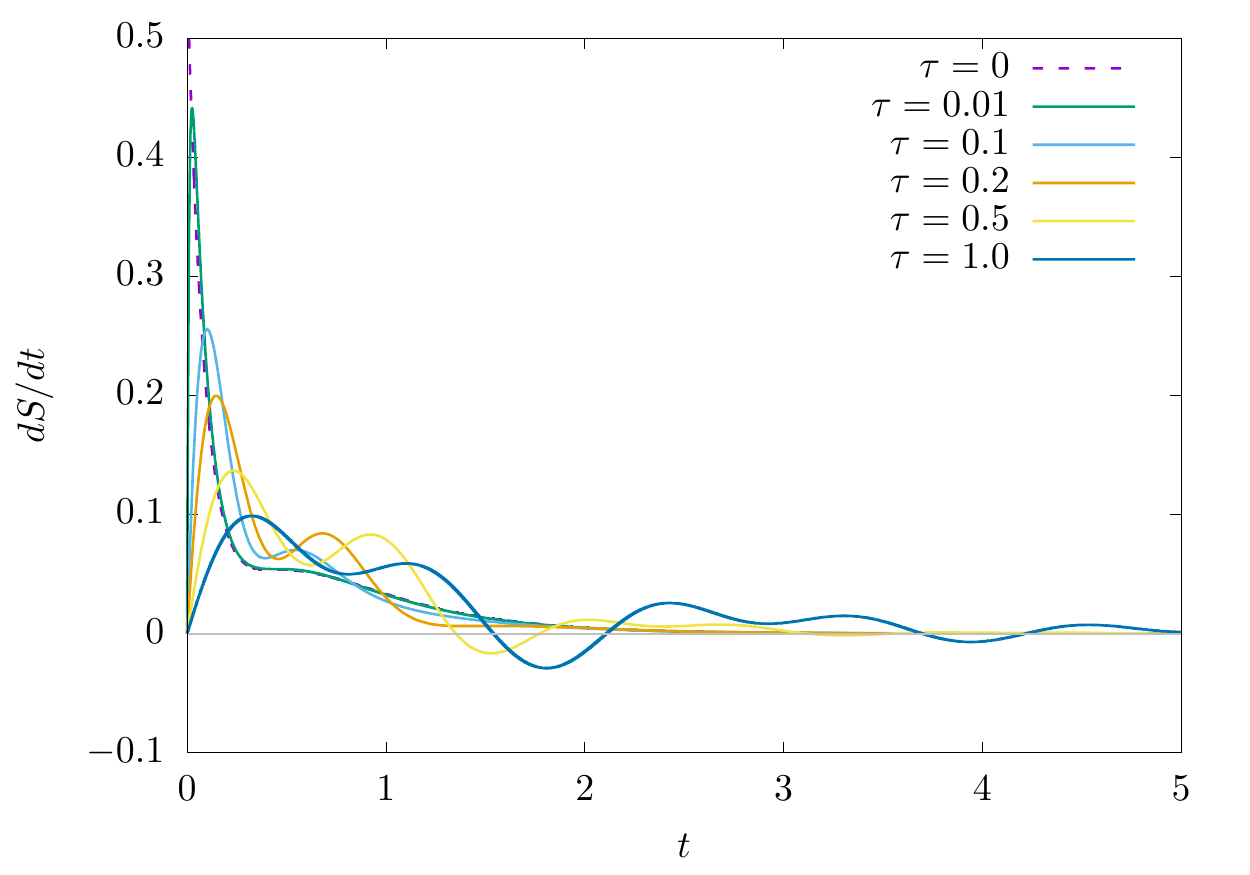}
\caption{The results above represent the time derivative of the entropy for $\tau=0, 0.01, 0.1, 0.2, 0.5, 1.0$. The other variables are $ \gamma = 5, m = 1 $. In this case $\tau_c= m/(4\gamma)= 0.05.$ Observe that for $\tau=0$ we have $dS/dt>0$, hence no backflow of entropy.}
\label{figdsdtversustau}
\end{center}
\end{figure}

The presence of oscillations due to $\kappa_I\neq 0$ does not guarantee information backflow. There are other factors that contribute to whether the entropy will decrease or not. In order to understand this point, let us take a look at Fig.~\ref{figdsdtversustau}. For instance, if $\tau=0$ then the noise becomes Gaussian white and the entropy will not decrease regardless of $\kappa_I\neq 0$, for the pulling protocol studied here. Information backflow is a direct  manifestation of the coupling of the non-Markovian memory kernel with a high dissipation regime. 

The phase diagrams in  figure~\ref{fig-discriminant} shows the ranges of parameters that favor oscillations in the entropy (in the sense of peaks on the spectrum, such as those in figure~\ref{entropyspectrumlinear}). However, in the high dissipation range, on the right in the diagrams of figure~\ref{fig-discriminant}, information backflow might as well happen. In order to check for it, we tested the existence of the backflow for several values below and above $\tau_c$, as shown in figure~\ref{figdsdtversustau}.  We observe that: for $\tau=0$ no backflow is present, as expected; for $\tau>\tau_c=0.05$ we observe several instances of backflow. However, for $\tau=0.2$ no backflow is observed. Thus, even for high dissipation values, or equivalently large values of $\tau$, observing the backflow is not guaranteed.

In the next section we include a small amount of non-linearity in our model via a weak quartic potential $k_3x^4/4$ (so that typically $k_3T/k_2^2\ll1$). In this we we expect to transcend some of the strange behaviors of harmonic systems, and on the other hand to be able to reasonably compare the results with those for the harmonic system. 

\section{Numerical results for the non-linear model}
In our simulations, we have run a series of runs of the process for the range of parameters given by $m=1.0,\, k_1 = 2.0, \,k_2=2.0,\,k_3=0.005,\,L_0=1.0,\, \, \tau = 1.0,\, T = 1.0,\,\, \gamma=0.5,\,\lambda=1.0.$   The number of runs of the protocol driven process is $2\times 10^{5}$. The results presented in the following correspond to averages over these simulations. The initial states are sampled over with the thermalized equilibrium distribution.

\subsection{Heat analysis}
We now generalize the harmonic model to a non-harmonic one by including a weak quartic potential term $k_3\,x^4/4$ into the interaction. The results of Section IV will no longer strictly apply, but they are a baseline that will be used for comparison sake. The chosen value for the potential terms obey~\cite{Defaveri2017,Defaveri2018} $k_3T/k_2^2\ll 1$, so the quartic potential can be considered as a small energetic correction for the harmonic potential energy.

\begin{figure}
\begin{center}
\includegraphics[height=3.0in,width=4.0in,angle=0]{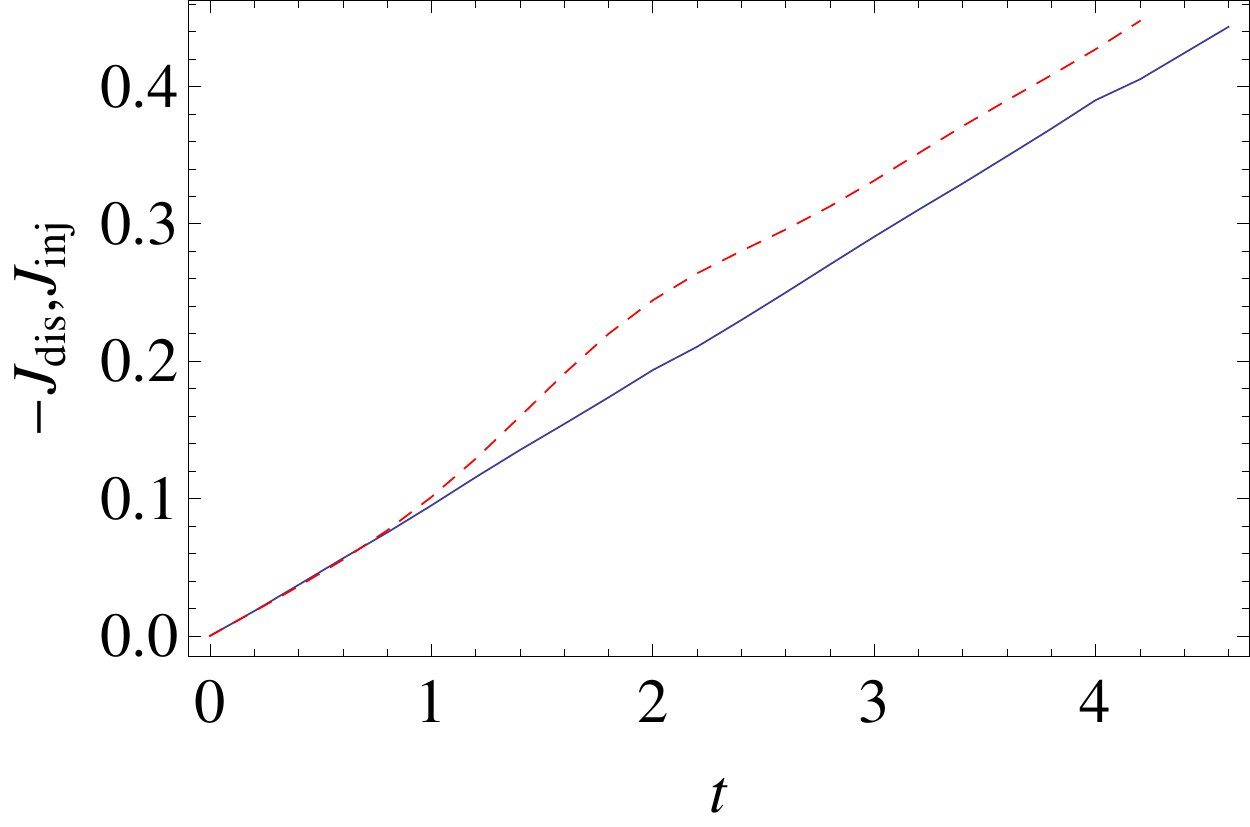}
\caption{Injected (dashed red line) and dissipated (continuous blue line) heat during the protocol.}\label{jinjjdiss}
\end{center}
\end{figure}

We can split the total heat exchanged with the reservoir up to time $t$, $Q(t)$, into the  injected  $J_{\rm inj}(t)$ and dissipated parts $J_{\rm diss}(t)$.m as defined below.
\begin{eqnarray}
 Q(t) &=& \Delta J(t), \nonumber\\ 
 &=& J_{\rm inj}(t)+J_{\rm diss}(t),
 \end{eqnarray}
where
\begin{eqnarray}
 J_{\rm inj}(t) &=& \int_{0}^{t}ds\,\xi(s)\,v(s),\\
 J_{\rm diss}(t) &=& - \int_{0}^{t}ds\,\int_{0}^{s}dt^{\prime}\, \phi(t-t^{\prime})\, v(t^{\prime})\, v(t).
\end{eqnarray}
In Fig.~\ref{jinjjdiss} we show the results from the simulations. 

\begin{figure}
\begin{center}
\includegraphics[height=3.0in,width=4.0in,angle=0]{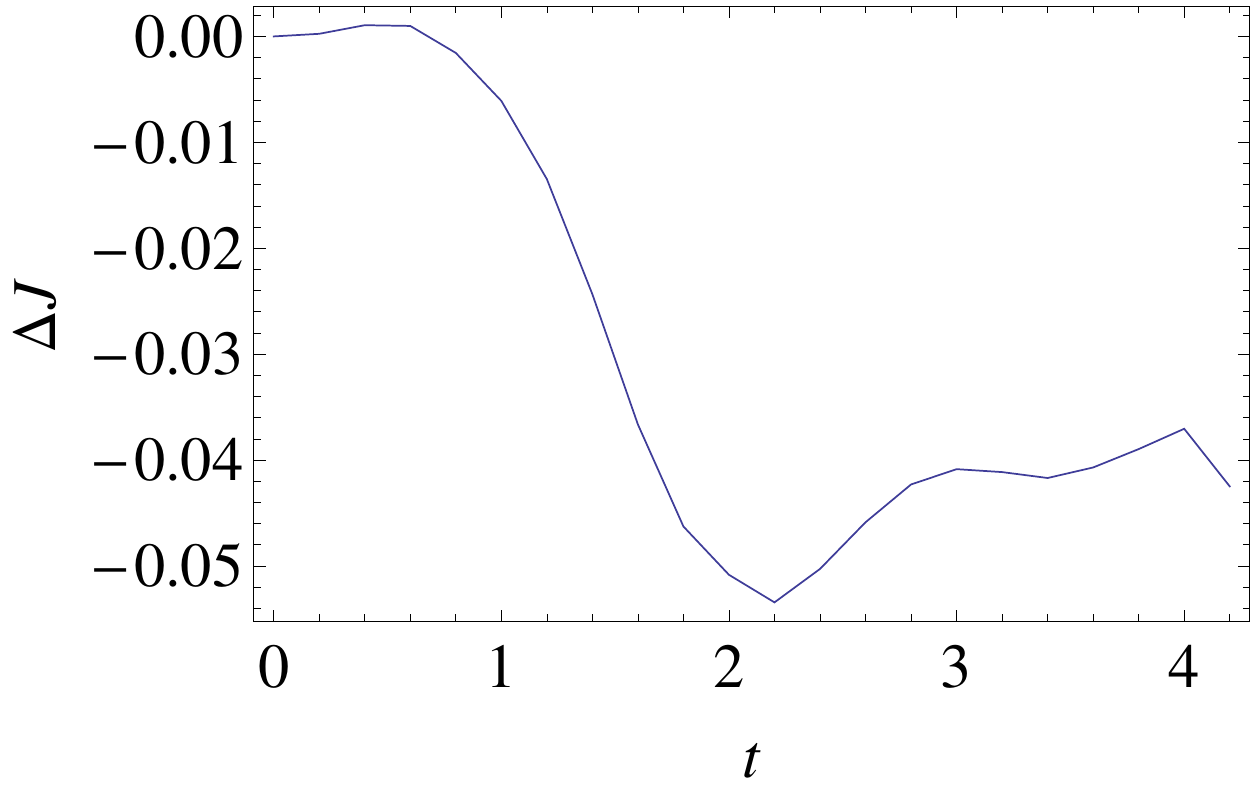}
\caption{Total heat absorbed by the system. Observe that it tends to a negative value since it expresses the dissipated work done by the external system during the protocol.}\label{DeltaJ}
\end{center}
\end{figure}

In Fig.~\ref{DeltaJ},  we learn the total heat exchanged $\Delta J(t)$ tends to saturate around a negative value. This is due to the fact that part of the work, done upon the system during the protocol, becomes heat and is transferred to the reservoir.

\begin{figure}
\begin{center}
\includegraphics[height=3.0in,width=4.0in,angle=0]{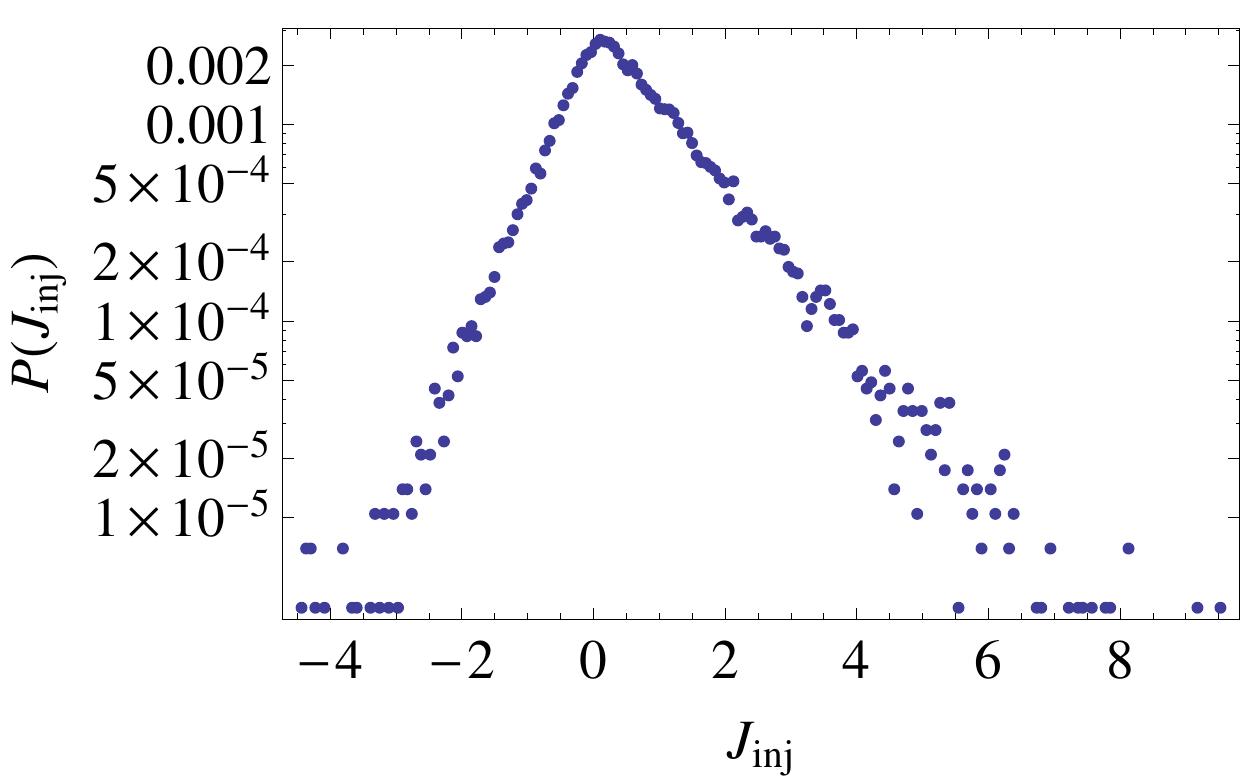}
\caption{The probability distribution for the injected heat $J_{\rm inj}$ for the complete protocol exhibits a markedly exponential behavior at the tails. It suggests that it satisfies a fluctuation theorem of sorts.}\label{pjinj}
\end{center}
\end{figure}

\begin{figure}
\begin{center}
\includegraphics[height=3.0in,width=4.0in,angle=0]{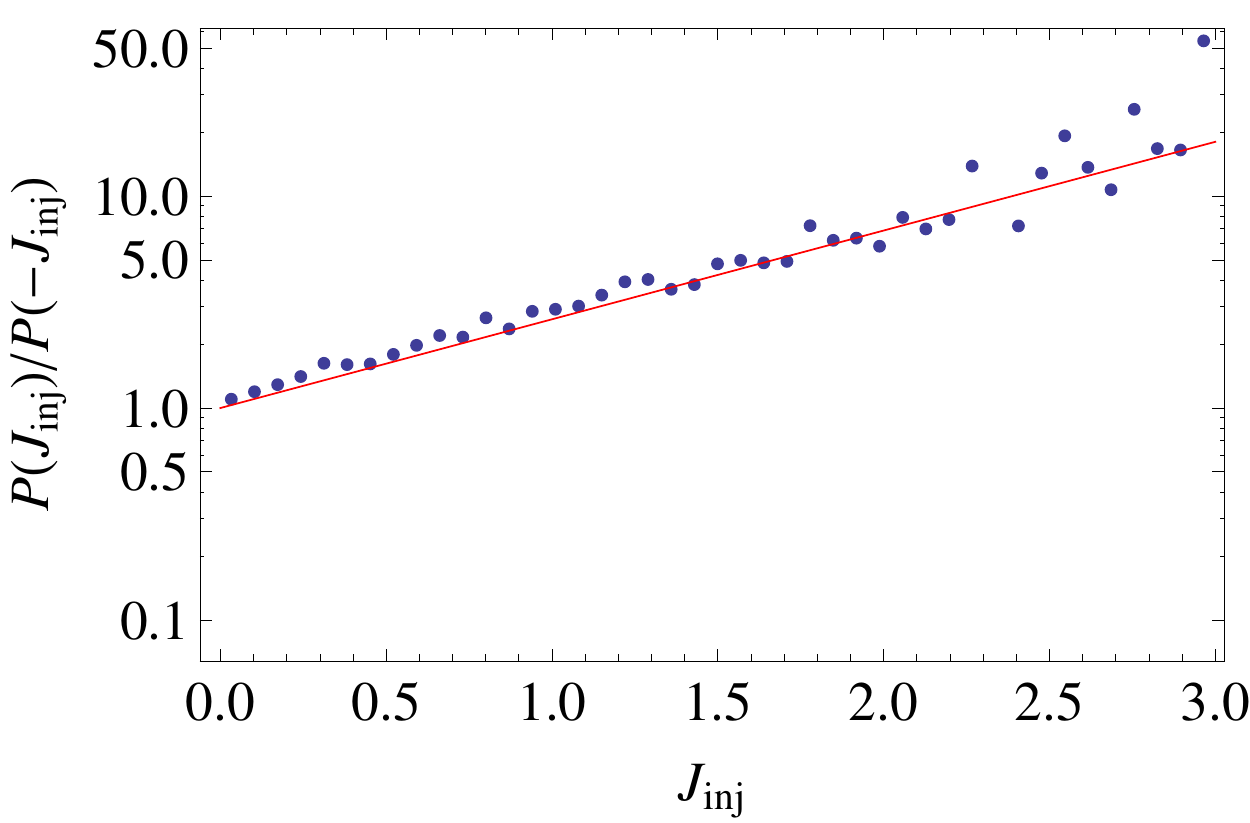}
\caption{The fluctuations off the injected heat for the whole protocol obey a fluctuation theorem form. The red line is not a data adjustment: it is the theoretical curve obtained in reference~\cite{Morgado2014} for a Markovian model, hence the slight non-conformity of the data points. The actual adjustment would have a higher angular coefficient, as a consequence of the non-Markovian memory kernel.}\label{frjinj}
\end{center}
\end{figure}

We can also obtain the probability distribution for the injected heat $p(J_{inj})$ for the duration of the protocol. This is shown in Fig.~\ref{pjinj}. It shows exponential tails and clearly suggests that a relations of the fluctuation theorem form~\cite{Morgado2014}
\begin{equation}
\ln \frac{p(J_{\rm inj})}{p(-J_{\rm inj})} = 2\frac{\mu _{J_{\rm inj}}}{\sigma ^2 _{J_{\rm inj}}} J_{inj},
\end{equation}
shall hold. Actually this is verified in Fig.~\ref{frjinj} to a very good degree.

That kind of behavior is already well known~\cite{Morgado2014,Morgado2016} where the distributions are obtained for systems in contact with thermal and athermal heat baths. The action of the pulling protocol in the system will be felt as an equivalent thermodynamic work transfer, as we see in the following.

\subsection{Work analysis}
\begin{figure}
\begin{center}
\includegraphics[height=3.0in,width=4.0in,angle=0]{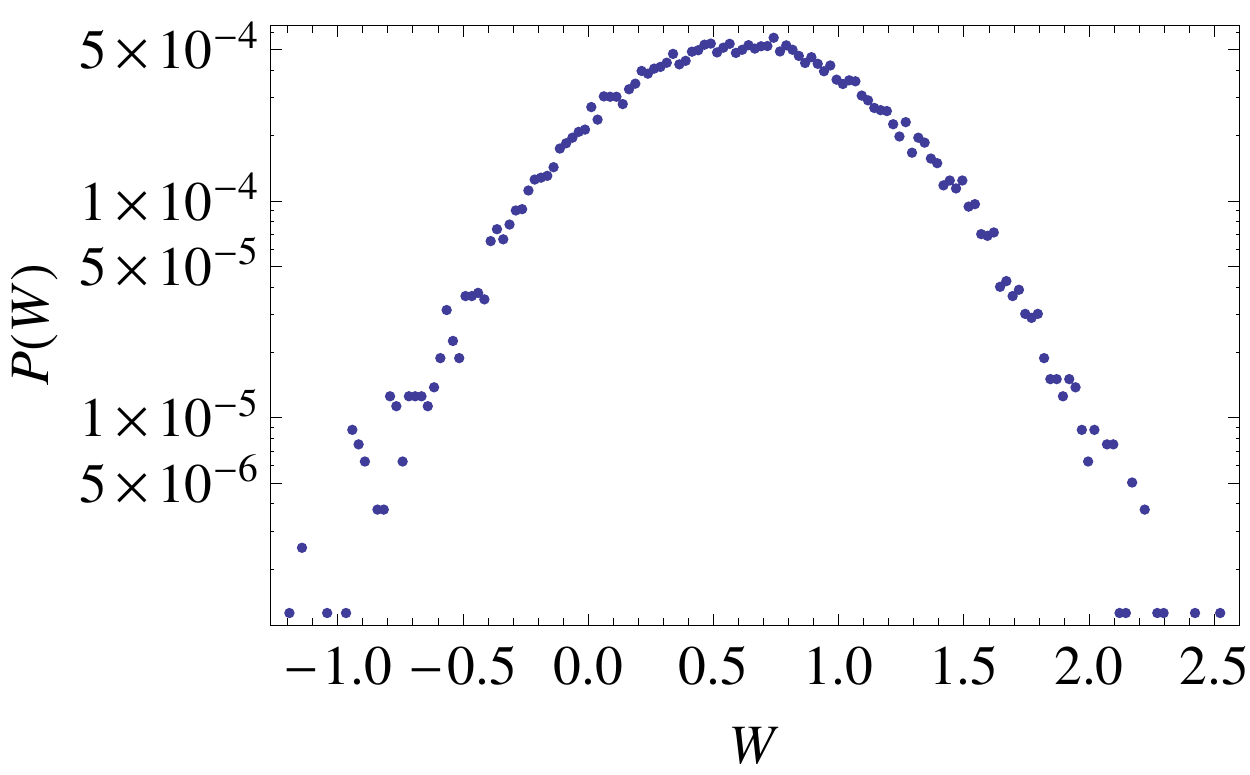}
\caption{Probability distribution for the work $W$ done upon the system by the external system.  The form of the work fits a Gaussian distribution, already found for similar models~\cite{Morgado2010}}\label{pwvw}
\end{center}
\end{figure}

The work probability has an symmetric form around a non-zero average (see Fig.~\ref{pwvw}, where it is clear that $\left<W_{ext}\right>>0$), displaced to the positive $W$ side), since the external work is done by stretching the spring ($L(0)=0\rightarrow L(t) >0 $). The distribution displayed in Fig.~\ref{pwvw} shows a Gaussian character
\begin{equation}
p\left( W \right) = \frac{1}{\sqrt{2 \pi \sigma ^2 _W}} 
\exp \left[ \frac{\left(W - \mu _W \right)^2}{2 \, \sigma ^2 _W} \right]
\end{equation}
The same dependence has been found for a similar model~\cite{Morgado2010} (on difference was $k_3=0$). A fluctuation relation can be extracted on the form
\begin{equation}
\ln \frac{p(W)}{p(-W)} = 2\frac{\mu _{W}}{\sigma ^2 _{W}} W,
\end{equation}
which agrees with the Gaussian character of the work distribution, and satisfies Crooks~\cite{Crooks2000} and Jarzynski~\cite{Jarzynski1997a} relations.

\begin{figure}
\begin{center}
\includegraphics[height=3.0in,width=4.0in,angle=0]{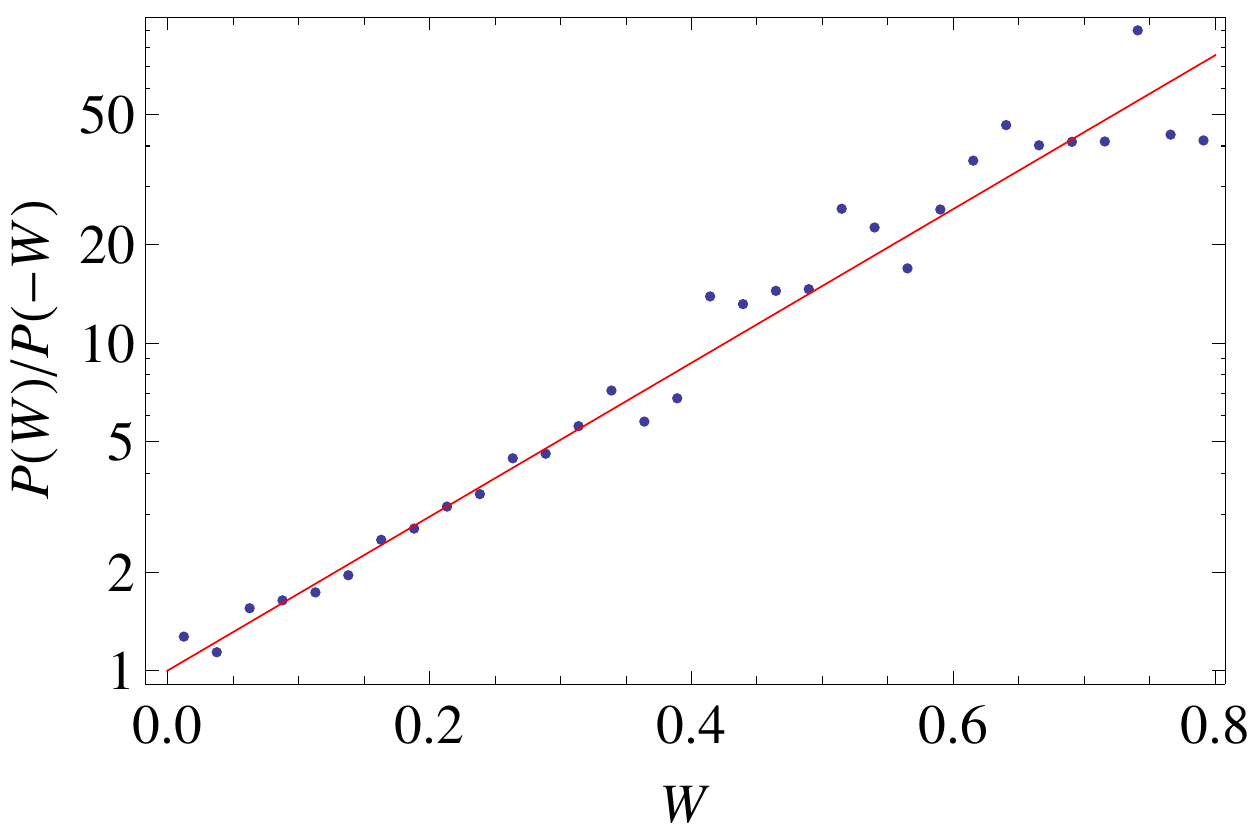}
\caption{Fluctuation relation obtained for the work transfer from the external system. Observe that the relation above upholds the Gaussian character of the work distribution.}\label{frwvw}
\end{center}
\end{figure}

\subsection{Entropy analysis}

\begin{figure}
\begin{center}
\includegraphics[height=3.0in,width=4.0in,angle=0]{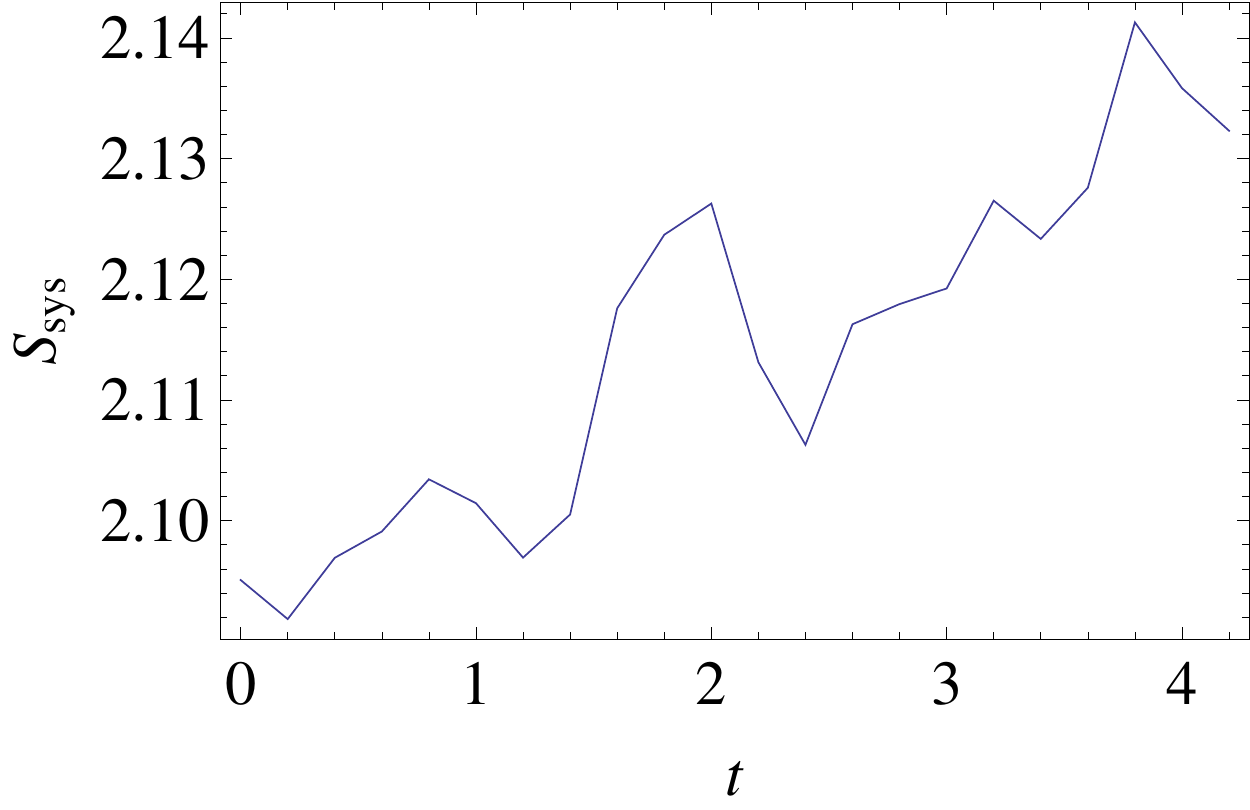}
\caption{Variation of the entropy as a function of time for the case of non-linear system under a Jarzynski pulling protocol.}\label{ssys}
\end{center}
\end{figure}

\begin{figure}
\begin{center}
\includegraphics[height=6.0in,width=4.0in,angle=0]{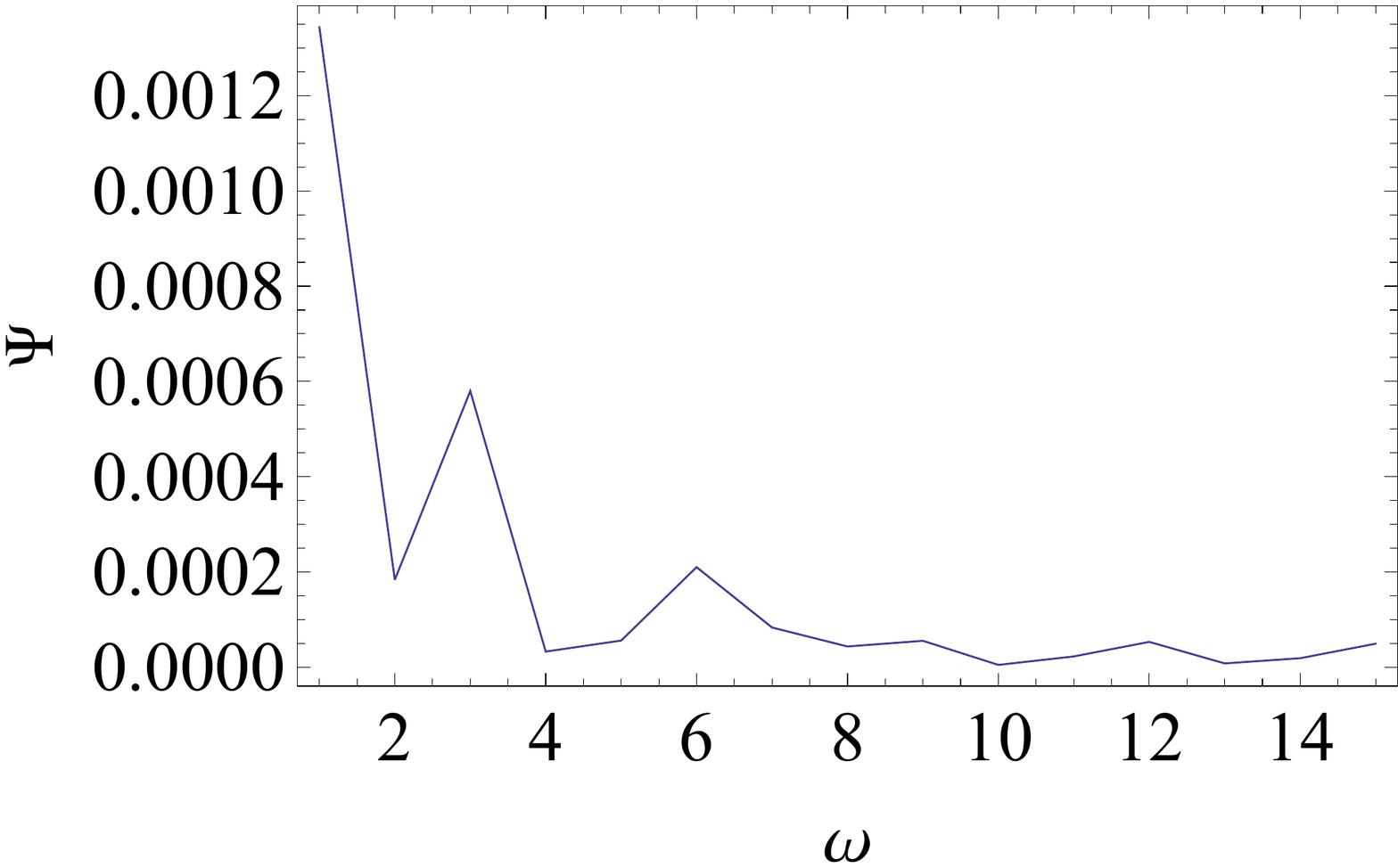}
\caption{We analyze the spectrum of the entropy variations. The peaks at, approximately, 3, 6, 9, 12 and 15 are clearly visible.}\label{powerspectrum}
\end{center}
\end{figure}

In accordance with the oscillatory behavior for the superposition of injected and dissipated heat (see Fig.~\ref{DeltaJ}), we analysed the behavior of the entropy of the system, $S_{\rm sys}(t)$, which shows oscillations and information backflow. The oscillatory dependence induced us to perform a spectral analysis which indicates a peak close to $\omega = 3$ as in Fig~\ref{powerspectrum}. Comparing the non-linear simulation shown in Fig.~\ref{ssys} with the linear model where all parameters are the same, except for $k_3=0$ was done. This is shown in figure~\ref{entrpoyfromzerotomax}. Interestingly, we noticed that the linear non-Markovian model  does not present information backflow while the non-linear model does.

\section{Concluding Remarks}
In the present work, we studied the energetics, and the entropic, aspects of non-Markovian massive models subjected to external pulling protocols, obeying the Jarzynski equality (JE). The importance of rare events for the non-equilibrium dynamics of a system Cannot be downplayed. The verification of the JE only occurs thanks to these rare events, as can be readily calculated for a few cases, such as the one presented herein.

More specifically, we work out two models, a linear (harmonic) one and a (slightly) non-linear (an-harmonic) one. The linear model allows for exact analytical treatments, while we exploit the non-linear case numerically. 

The non-Markovian harmonic model can be solved exactly, and we can obtain exact  probability  distribution functions for its dynamic variables. Harmonic models can exhibit quite singular behavior in the context of small  classical system thermodynamic behavior. For instance, such models lead to ballistic heat conduction not obeying Fourier Law; or that strictly harmonic potentials (with time invariant spring hardness) cannot be used to build machines with positive efficiencies.  

We have first demonstrated, exactly, the JE for a class of protocols that are in fact quite general. We also have studied the mechanical effects of the non-Markovian memory kernel. Its highly unusual properties can be appreciated by focusing in the behavior of the instantaneous entropy.

The evolution of the Shannon informational entropy for the harmonic system can be studied exactly, since the initial state corresponds is described by a Gaussian equilibrium distribution. The Gaussian character of the noise, and the linearity between variables and noise, guarantees that the probability distribution for the harmonic non-Markovian system is always Gaussian, although not of a Boltzmann-Gibbs format, hence a non-equilibrium distribution.

For the harmonic non-Markovian model, the time evolution of the informational entropy can be obtained exactly, and it does not depend on the protocol at all. In fact, for the quasi-static protocol, a system in contact with a reservoir at temperature $T$, starting at an equilibrium state at temperature $T$, would always be at equilibrium, the entropy would not vary. Two interesting cases may happen: firstly, if the system is initially at equilibrium, at the same temperature of the reservoir, it stays at equilibrium, regardless of the protocol. The only action of the protocol is to change the average position of the Brownian particle during the process; Secondly, if the system starts in equilibrium, at a temperature which is distinct from the reservoir's, the system will reach a non-equilibrium state where all the entropy variation is due to flux to and fro the reservoir, since no entropy is produced by the action of the external work protocol no matter how apparently far from equilibrium are its actions!
The effect above is one more strange consequence of the harmonic type of potentials.

The presence of the non-Markovian memory kernel may induce actual oscillations on the entropy. Akin to the velocity oscillations, the  entropy oscillations are due to the fact that the memory kernel time-scale $\tau \neq 0.$ It disappears as $\tau\rightarrow0.$ Thus, for the appropriate range of parameters, the memory kernel acts as an information pump, recovering it (partially) from the thermal bath. In fact, this constitutes strong evidence that the presence of the memory kernel indicates the formation of structures (which can store information) in the bath, such as slow hydrodynamic modes in Brownian-Fluid models.    
Taking a more realist approach to the problem, we studied a non-harmonic model, where we introduced a small quartic potential as a perturbation term.  Similarly to earlier models, the injected heat yields a fluctuation relation in the form of an asymmetric large deviation function. The work transmitted from the external system obeys the Crooks relation.

The presence of the non-linear terms somehow restores ``normality'' to the evolution of the total entropy and its production during the protocol. In this case we observe that the entropy production rate, by the system, is non-zero. Concerning the entropy oscillations, they are persistent since their cause is that $\tau\neq0$ and  distinct harmonics can be detected by spectral analysis. The principal components are the same as the harmonic case if $k_3/k_1^2\ll1$.

\begin{center}
    {\bf Acknowledgements}
\end{center} 
W.A.M.M. would like to thank the Brazilian agency CNPq. D.O.S.P. acknowledges the Brazilian funding agencies CNPq (Grants No. 307028/2019-4), FAPESP (Grant No. 2017/03727-0) and the Brazilian National Institute of Science and Technology of Quantum Information (INCT/IQ). This study was financed in part by Coordena\c c\~ ao de Aperfei\c coamento de Pessoal de N\' ivel Superior - Brasil (CAPES) - Finance Code 001.




\appendix

\section{A mechanical view of the non-Markovian kernel}\label{appA}

\begin{figure}
	\includegraphics[scale=1]{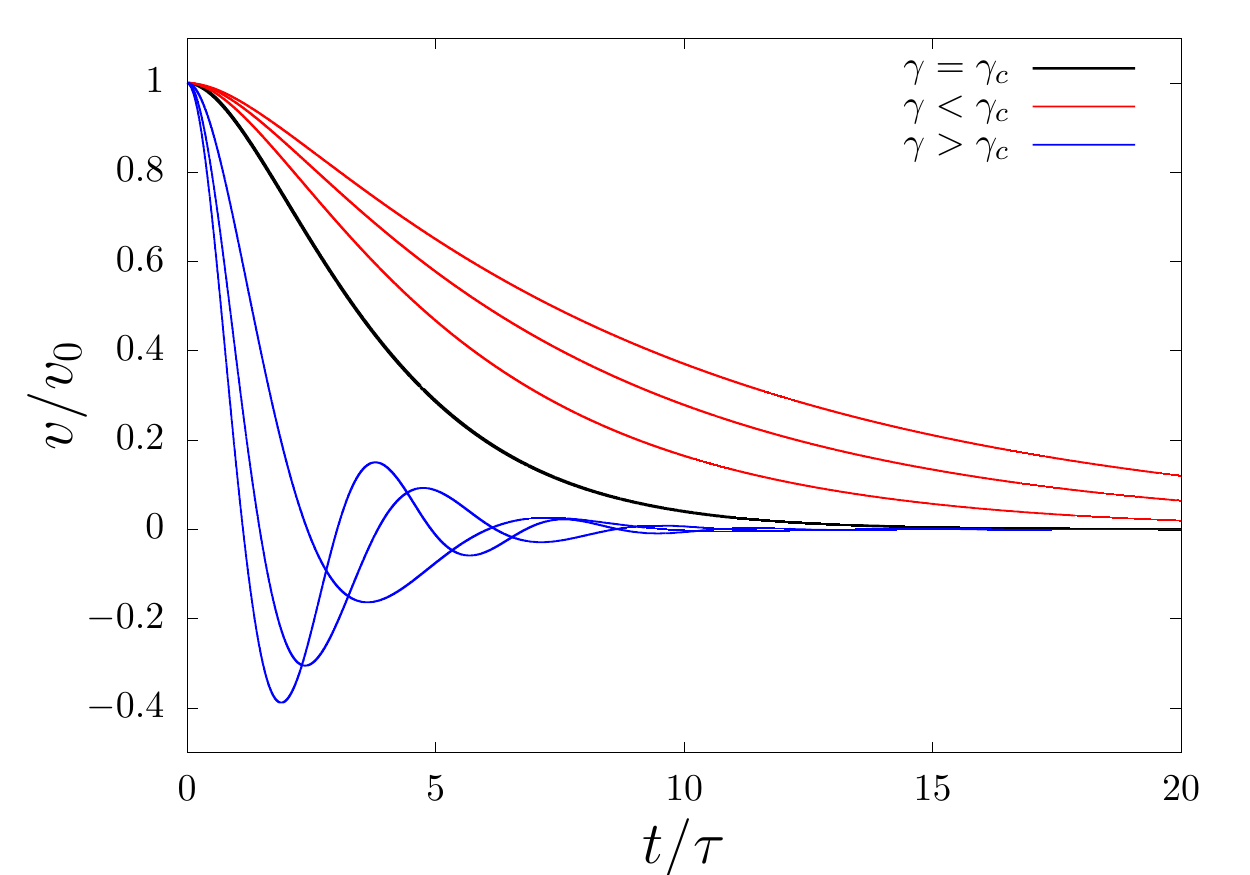}
	\caption{We highlight the behavious of the three regimes.}\label{figvv_0nonmarkov}
\end{figure}

As an illustration of the behavior induced by the memory kernel, let us study the velocity attenuation when the interaction potentials and the energy injection are turned off. Thus, starting from the simplified equation of motion, the non-Markovian dissipation dynamics reads
\begin{eqnarray}
m \dot{v} + \int_0^t \phi(t-t') v(t') dt' = m \dot{v} + \int_0^t \frac{\gamma}{\tau} e^{-(t-t')/\tau} v(t') dt' = 0,
\end{eqnarray}
with initial condition $v_0 \neq 0$. To simplify the problem, let us define the inverse of the dissipation time-scale $\tau_{diss}^{-1} = \Gamma = \gamma/m$ and let us re-scale time by $\tau$ (effectively making $\tau=1$) so that the equation of motion may be written in the far simpler form:
\begin{eqnarray}
\dot{v} + \Gamma \int_0^t e^{t-t'} v(t') dt' = 0. 
\end{eqnarray}
To solve this equation we will employ the Laplace transform, we obtain
\begin{eqnarray}
s \tilde{v}(s) - v_0 + \Gamma \frac{\tilde{v}(s)}{1 + s} = 0 \longrightarrow  \tilde{v}(s) = \frac{ (1+s)}{s^2 + s + \Gamma} v_0.
\end{eqnarray}
The inverse can be calculated by using a version of Mielin integration (rotated by $\pi$/2 in the complex plane) as
\begin{eqnarray}
v(t) = v_0 \int_{-\infty}^\infty \frac{dq}{2 \pi} \frac{1 + i q + \epsilon}{(i q + \epsilon)^2 + (i q + \epsilon) + \Gamma} e^{i q t}.
\end{eqnarray}
The integration is evaluated using the residue theorem, so we are interested in the poles from the roots of the denominator:
\begin{eqnarray}
q_{\pm} = \frac{i}{2} \pm \sqrt{\Gamma - \frac{1}{4}}.
\end{eqnarray}
We can see that, depending on the value of $\Gamma$, we may have three distinct regimes (depicted in Fig.~\ref{figvv_0nonmarkov}):
\begin{eqnarray}
\frac{v(t)}{v_0} = \left\{ \begin{array}{l}
e^{-t/2} \left\{ \cosh \left( t \sqrt{\frac{1}{4} - \Gamma} \right) + \frac{\sinh \left( t \sqrt{\frac{1}{4} - \Gamma} \right)}{\sqrt{1 - 4 \Gamma}}  \right\} ~~~ \big( \Gamma < \frac{1}{4} \big), \\
e^{-t/2} \left(1 + \frac{t}{2}\right) ~~~~~~~~~~~~~~~~~~~~~~~~~~~~~~~~~~~~~~~ \big( \Gamma = \frac{1}{4} \big), \\
e^{-t/2} \left\{ \cos \left( t \sqrt{ \Gamma - \frac{1}{4}} \right) + \frac{\sin \left(  t \sqrt{ \Gamma - \frac{1}{4} } \right)}{\sqrt{4 \Gamma-1}}  \right\} ~~~~~~~~~ \big( \Gamma > \frac{1}{4} \big).
\end{array} \right.
\end{eqnarray}
For a similar calculation, see reference~\cite{Bouchaud1998}.

Remarkably, only for sufficiently large values of $\Gamma$ we obtain {\bf oscillations}, e.g.
\[
\frac{\tau_{diss}}{\tau} < 4.
\]
Expressing this result in the original variables, the critical damping is
\begin{eqnarray}
\frac{4 \tau \gamma_c}{m} = 1. \label{eq-restriction_mechanical}
\end{eqnarray}

For the high dissipation regime, for the situations when the velocity of the particle reaches zero, the memory function brings back information from the past motion of the particle accelerating it back to non-zero velocities (of opposite sign, of course).

\section{Behaviour of $\kappa_i$}
\label{appendix-discriminant}
In order to understand the nature of the Green function $g(t)$ and the auxiliary function $f(t)$ defined in section \ref{section-entropy_calculation}, we must study the nature of the roots of the numerator of $R(s)$, which we will refer as $\kappa_i$, that is
\begin{eqnarray}
m \tau (s - \kappa_1)(s-\kappa_2)(s-\kappa_3) = m \tau s^3 + m s^2 + ( \Gamma + \tau \nu^2 ) s + \nu^2,
\end{eqnarray}
where we have used $\Gamma = \gamma/m$ and $\nu^2 = (k_1+k_2)/m$.
We can use the discriminant of the numerator of $R(s)$ that we shall refer as $\Delta$ and is calculated as
\begin{eqnarray}
 \Delta = \Gamma^2 - 4 \Gamma^3 \tau - 4 \nu^2 + 4 \Gamma \tau (5 - 3 \Gamma \tau) \nu^2 - 4 \tau^2 (2 + 3 \Gamma \tau)\nu^4 - 4 \tau^4 \nu^6.
\end{eqnarray}
Note that if we make $\tau = 0$ the discriminant becomes $\Delta = \Gamma^2 - 4 \nu^2$, which is the usual result for a system with white noise.

\begin{figure}
	\includegraphics[width=0.48\textwidth]{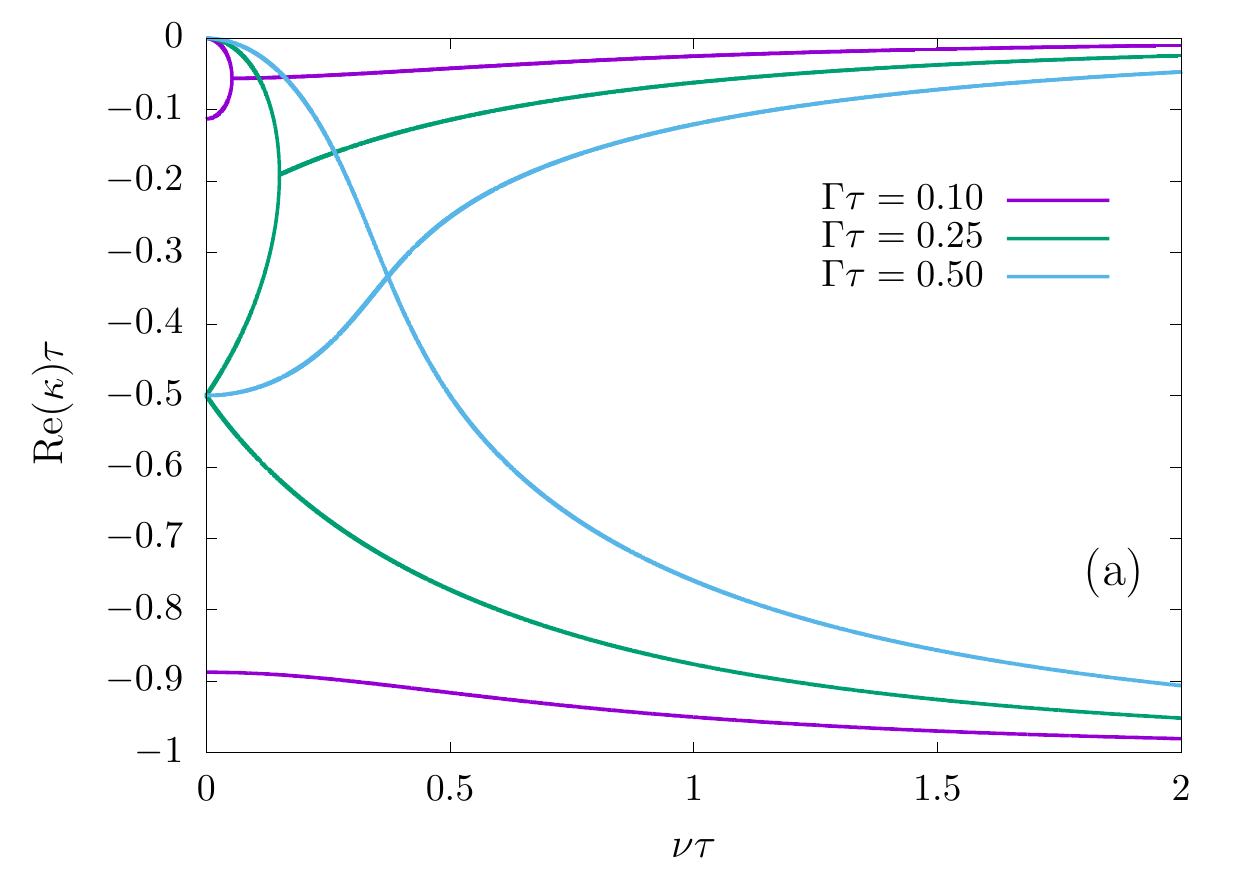}
	\includegraphics[width=0.48\textwidth]{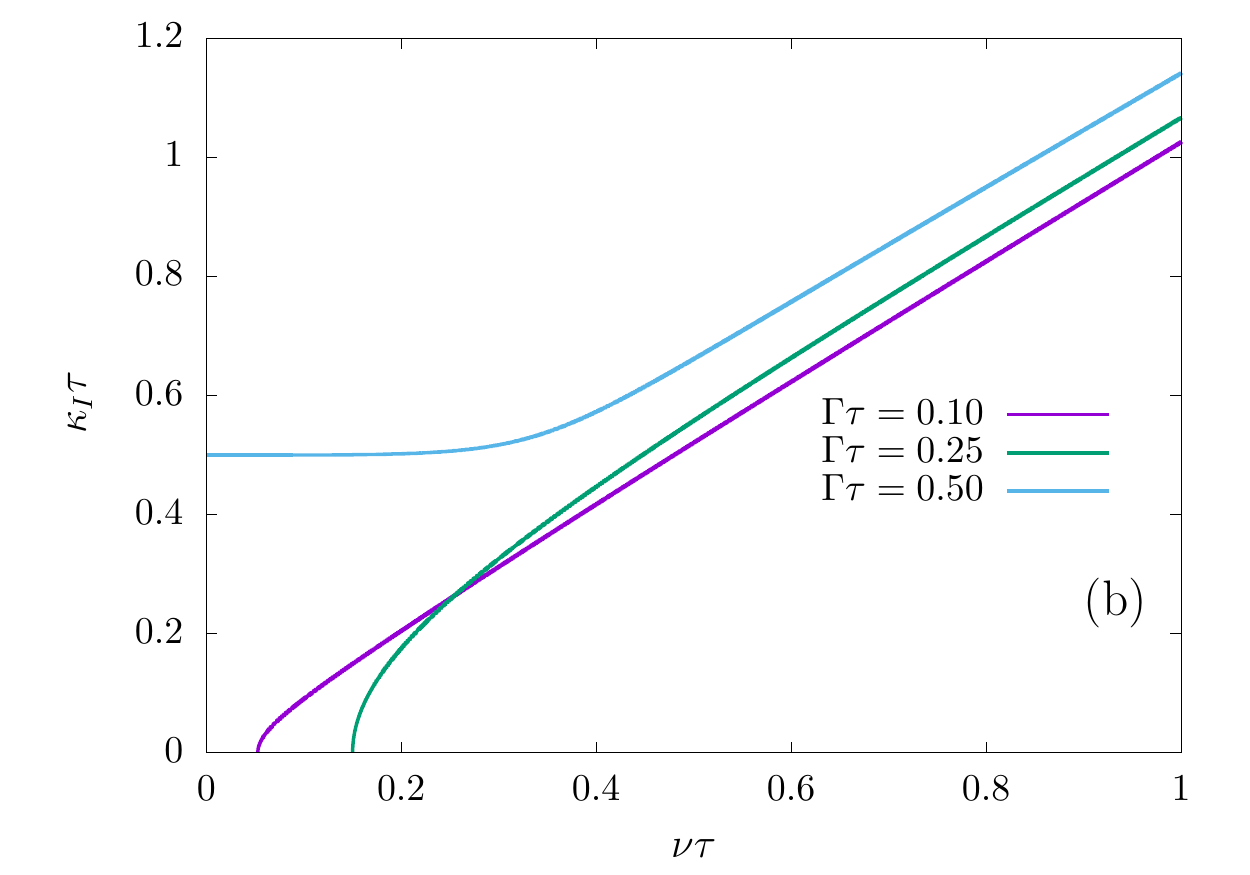}
	\caption{We show the real part of the roots $\kappa$ in panel (a) and the imaginary part of the roots $\kappa$ (labeled $\kappa_I$) in panel (b) as a function of $\nu$ for different values of $\Gamma$ (all scaled by $\tau$). It is possible to note that the real part is always negative and that $\kappa_I$ grows with $\nu$ for sufficiently large values. }\label{fig-roots}
\end{figure}

It is possible to obtain the values of $\kappa$'s exactly by solving the third degree polynomial. The expressions become:
\begin{eqnarray}
 \kappa_1 &=& - \frac{1}{3 \tau} \left\{ 1 - \frac{2 - 6 \Gamma \tau - 6 \Gamma^2 \tau^2 + 2^{1/3} \left( - 2 + 9 \Gamma \tau - 18 \tau^2 \nu^2 + \sqrt{- 27 \tau^2 \Delta } \right)^{2/3}}{ 2^{2/3} \left( - 2 + 9 \Gamma \tau - 18 \tau^2 \nu^2 + \sqrt{- 27 \tau^2 \Delta } \right)^{1/3}} \right\} \nonumber \\
  \kappa_2 &=& - \frac{1}{3 \tau} \left\{ 1 + \frac{2 - 6 \Gamma \tau - 6 \Gamma^2 \tau^2 -(-2)^{1/3}  \left( - 2 + 9 \Gamma \tau - 18 \tau^2 \nu^2 + \sqrt{- 27 \tau^2 \Delta } \right)^{2/3}}{ 2^{2/3} \left(  2 - 9 \Gamma \tau + 18 \tau^2 \nu^2 - \sqrt{- 27 \tau^2 \Delta } \right)^{1/3}} \right\} \\ 
  \kappa_3 &=& - \frac{1}{3 \tau} \left\{ 1 - (-1)^{2/3} \frac{2 - \Gamma \tau - 6 \Gamma^2 \tau^2 + (-1)^{2/3}(2)^{1/3} \left( - 2 + 9 \Gamma \tau - 18 \tau^2 \nu^2 + \sqrt{- 27 \tau^2 \Delta } \right)^{2/3}}{ 2^{2/3} \left( - 2 + 9 \Gamma \tau - 18 \tau^2 \nu^2 + \sqrt{- 27 \tau^2 \Delta } \right)^{1/3}} \right\}, \nonumber
\end{eqnarray}
where to simplify the answer we used $\Delta$ as the discriminant. Note that the dependence between the variables ($\Gamma$, $\nu$ and $\tau$) is highly nontrivial, and from the solutions is not very clear the regimes one could obtain. We demonstrate some typical values for a couple of examples in Fig. \ref{fig-roots}.

Despite the complexity, some general properties can still be extracted. The real component of the $\kappa$'s will always be negative, indicating that the solutions will always approach a limit, and never diverge. Since all the coefficients of the polynomial are real and positive, the sign of the discriminant will determine the nature of the roots. We are interested in differentiating the cases where all roots are real ($\Delta \geq 0$), and the oscillating case where two roots are complex conjugate of each other ($\Delta < 0$).

\begin{figure}
	\includegraphics[width=0.48\textwidth]{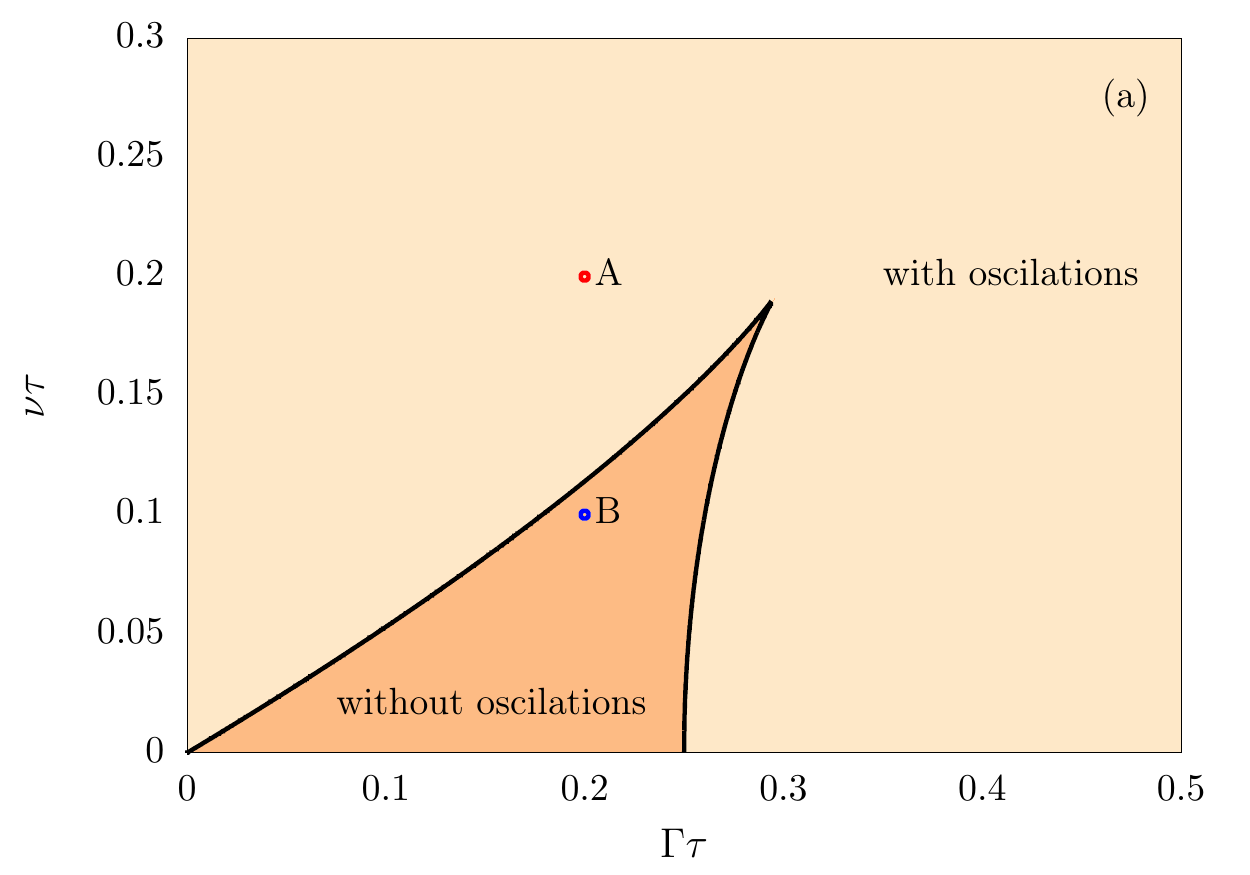}
	\includegraphics[width=0.48\textwidth]{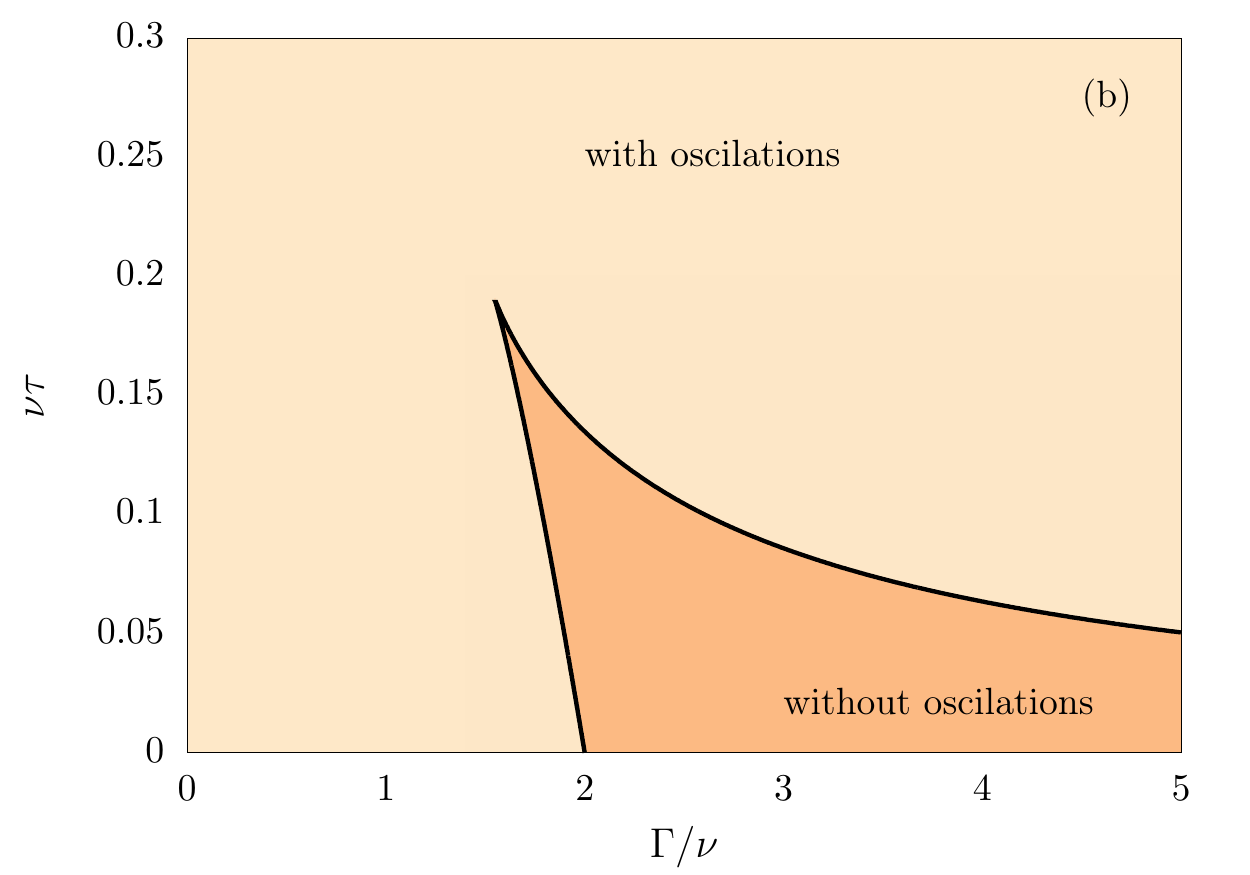}
	\caption{We highlight range of values for the system parameters so that oscillations may be observed. In each panel we use a different set of dimensionless parameters, for (a) we use $\nu \tau$ and $\Gamma \tau$ and for (b) we use $\nu \tau$ and $\gamma / \nu$. The two marked points will be studied in the next figure. Note that in panel (a), by making $\nu=0$ (no external force) we recover the result from equation (\ref{eq-restriction_mechanical}), that is $\Gamma_C > \gamma_C \tau/m = 1/4$, so that the system displays oscillations. And in panel (b), by making $\tau=0$ (removing the bath memory) we recover $\gamma/m > \nu$ which defines the underdamped of a harmonic oscillator. It is also important to note that for a sufficiently large value of $\tau$, the discriminant will always be negative and the system will display oscillations. }\label{fig-discriminant}
\end{figure}

For that end, we create a portrait that encompasses all possible signs of $\Delta$ by reducing to two parameters either by choosing $\tau$ as the timescale and using the dimensionless parameters $\nu \tau$ and $\Gamma \tau$ or choosing $1/\nu$ as the timescale and using the dimensionless parameters $\nu \tau$ and $\Gamma / \nu$. Both cases are displayed in Fig. \ref{fig-discriminant}.

\begin{figure}
	\includegraphics[scale=1]{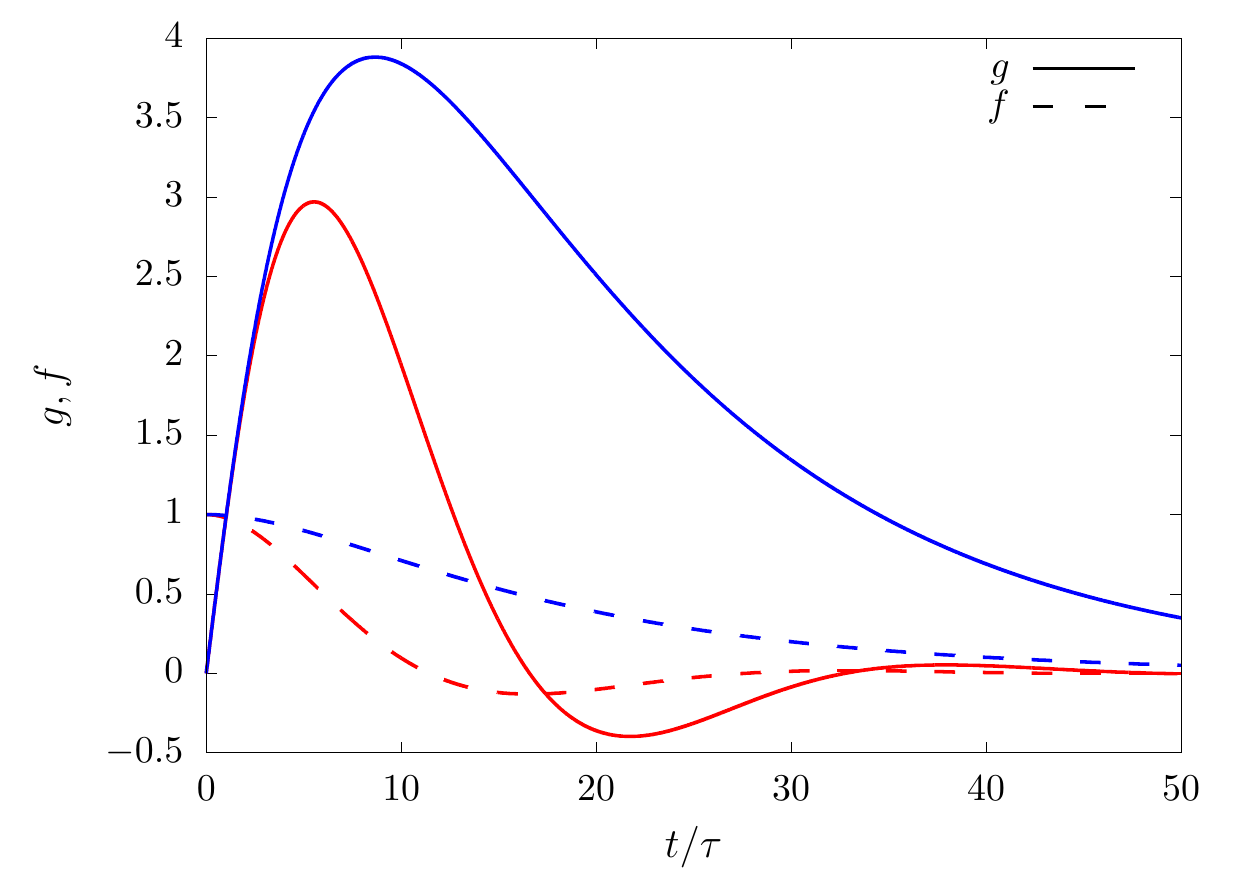}
	\caption{We highlight the behaviour of the Green function $g(t)$ and the auxiliary function $f(t)$ for the points A and B highlighted in Fig. \ref{fig-discriminant}. For a positive discriminant (the blue lines) the functions only decay while for a negative discriminant (the red lines) the functions display oscillatory behaviour. \label{fig-caseAB} }
\end{figure}

We can write the Green function, with a positive discriminant, as
\begin{eqnarray}
 g(t) = \frac{\kappa_1 \tau - 1}{m\tau (\kappa_1 - \kappa_2)(\kappa_1 - \kappa_3)}e^{- \kappa_1 t} + \frac{\kappa_2 \tau - 1}{m \tau (\kappa_2 - \kappa_1)(\kappa_2 - \kappa_3)}e^{- \kappa_2 t} + \frac{\kappa_3 \tau - 1}{m \tau (\kappa_3 - \kappa_1)(\kappa_3 - \kappa_2)}e^{- \kappa_3 t},
\end{eqnarray}
and for a negative discriminant ($\kappa_{2,3} = \kappa_R + \pm \mathrm{i} \kappa_I$) we have oscillations with frequency $\mathrm{Im}(\kappa_2) = \mathrm{Im}(\kappa_3) = \kappa_I$
\begin{eqnarray}
 g(t) = \frac{(1 - \kappa_1 \tau)e^{- \kappa_1 t}}{m \tau (\kappa_I^2 + (\kappa_1 - \kappa_R)^2)} + \frac{(\kappa_1 \tau - 1)e^{- \kappa_R t}}{m \tau (\kappa_I^2 + (\kappa_1 - \kappa_R)^2)}\cos(\kappa_I t) + \frac{(\kappa_1 - \kappa_R - \kappa_1 \kappa_R \tau + (\kappa_I^2 + \kappa_R^2)\tau)e^{-\kappa_R t}}{m \tau (\kappa_I^2 + (\kappa_1 - \kappa_R)^2)} \sin(\kappa_I t) ,
\end{eqnarray}
and the auxiliary function for positive discriminant is
\begin{eqnarray}
 f(t) = \frac{\kappa_1^2 \tau - \kappa_1 + \Gamma}{m \tau(\kappa_1 - \kappa_2)(\kappa_1 - \kappa_3)}e^{- \kappa_1 t} + \frac{\kappa_2^2 \tau - \kappa_2 + \Gamma}{m \tau (\kappa_2 - \kappa_1)(\kappa_2 - \kappa_3)}e^{- \kappa_2 t} + \frac{\kappa_3^2 \tau - \kappa_3 + \Gamma}{m \tau (\kappa_3 - \kappa_1)(\kappa_3 - \kappa_2)}e^{- \kappa_3 t},
\end{eqnarray}
and for negative discriminant
\begin{eqnarray}
 f(t) &=& \frac{\kappa_1^2 \tau - \kappa_1 + \Gamma}{m \tau (\kappa_1 - \kappa_2)(\kappa_1 - \kappa_3)}e^{- \kappa_1 t} + \frac{(\Gamma + \kappa_1 + (\kappa_I + \kappa_R(\kappa_T - 2 \kappa_1)\tau))e^{- \kappa_R t}}{m \tau (\kappa_I^2 + (\kappa_1 - \kappa_R)^2)}\cos(\kappa_I t) + \nonumber \\
 &+& \frac{(\Gamma + \kappa_1 + (\kappa_I^2 + \kappa_R(\kappa_R - 2 \kappa_1))\tau)e^{-\kappa_R t}}{m \tau (\kappa_I^2 + (\kappa_1 - \kappa_R)^2)} \sin(\kappa_I t).
\end{eqnarray}
In Fig. \ref{fig-caseAB} we display the behaviour of both regimes.

\section{Second order terms}
\label{appendix-As}
Here we present the expressions for the second order terms:
\begin{eqnarray*}
\mathcal{A}_1 &=& \left( {{\rm e}^{-\,{\frac{2\,\theta}{\lambda}}}}-2\,{{\rm e}^{{\frac 
{ \left( \kappa_{{1}}\lambda-1 \right) \theta}{\lambda}}}}+1 \right) \frac{T
\,\gamma\,{L_{{0}}}^{2}{k_{{2}}}^{2}}{{m}^{2}{\tau}^{2}{\kappa_{{1}}}
 \left( {\kappa_{{1}}}^{2}{\lambda}^{2}-1 \right) \left( {\kappa
_{{1}}}^{2}-{\kappa_{{2}}}^{2} \right) \left( {\kappa_{{1}}}^{2}
-{\kappa_{{3}}}^{2} \right)}\\ 
&-&  \left( {{\rm e}^{-\,{\frac {2\,\theta
}{\lambda}}}}-2\,{{\rm e}^{{\frac { \left( \kappa_{{2}}\lambda-1
 \right) \theta}{\lambda}}}}+1 \right) \frac{T\,\gamma\,{L_{{0}}}^{2}{k_{{2}}}^{2}}{{m}
^{2}{\tau}^{2}{\kappa_{{2}}} \left( {\kappa_{{2}}}^{2}{\lambda}
^{2}-1 \right) \left( {\kappa_{{1}}}^{2}-{\kappa_{{2}}}^{2}
 \right) \left( {\kappa_{{2}}}^{2}-{\kappa_{{3}}}^{2} \right)}\\
&+& \left( {{\rm e}^{-\,{\frac {2\,\theta}{\lambda}}}}-2\,{{\rm e}^{{
\frac { \left( \kappa_{{3}}\lambda-1 \right) \theta}{\lambda}}}}+1
 \right) \frac{T\,\gamma\,{L_{{0}}}^{2}{k_{{2}}}^{2}}{{m}^{2}{\tau}^{2}{\kappa_{{3}}} \left( {\kappa_{{3}}}^{2}{\lambda}^{2}-1 \right) \left( {
\kappa_{{1}}}^{2}-{\kappa_{{3}}}^{2} \right) \left( {\kappa_{{2}
}}^{2}-{\kappa_{{3}}}^{2} \right)}\\ 
&-& \left( {{\rm e}^{-\,{\frac {2\,
\theta}{\lambda}}}}-1 \right) \frac{T\,\gamma\,{L_{{0}}}^{2}{k_{{2}}}^{2}{\lambda}^{5
}}{{m}^{2}{\tau}^{2} \left( {\kappa_{{1}}}^{2}{\lambda}^{2}-1 \right)  \left( {\kappa_{{2}}}^{2}{\lambda}^{2}-1 \right) \left( {
\kappa_{{3}}}^{2}{\lambda}^{2}-1 \right)}\\ 
&+& \left( {{\rm e}^{{\frac { \left( \kappa_{{1}}\lambda-1 \right)
\theta}{\lambda}}}}-1 \right) ^{2} \frac{T\,\gamma\,{L_{{0}}}^
{2}{k_{{2}}}^{2}\left( 1+\tau\,\kappa_{{1}}
 \right)}{{m}^{2}{\tau}^{2} \left( \kappa_{{1}}\lambda-1 \right) ^{2
} \left( \kappa_{{1}}-\kappa_{{2}} \right) ^{2} \left( \kappa_{{1}}-
\kappa_{{3}} \right) ^{2}{\kappa_{{1}}}}\\ 
&+& \left( {
{\rm e}^{{\frac { \left( \kappa_{{2}}\lambda-1 \right) \theta}{\lambda
}}}}-1 \right) ^{2} \frac{T\,\gamma\,{L_{{0}}}^
{2}{k_{{2}}}^{2}\left( 1+\tau\,\kappa_{{2}} \right)}{{m}^{2}{\tau}
^{2} \left( \kappa_{{2}}\lambda-1 \right) ^{2} \left( \kappa_{{1}}-
\kappa_{{2}} \right) ^{2} \left( \kappa_{{2}}-\kappa_{{3}} \right) ^{
2}{\kappa_{{2}}}} \\
&+&\left( {{\rm e}^{{\frac { \left( \kappa_{{3}}\lambda-1
 \right) \theta}{\lambda}}}}-1 \right) ^{2} \frac{T\,\gamma\,{L_{{0}}}^
{2}{k_{{2}}}^{2} \left( 1+\tau\,\kappa_{{3}
} \right)} {{m}^{2}{\tau}^{2} \left( \kappa_{{3}}\lambda-1 \right) ^{
2} \left( \kappa_{{1}}-\kappa_{{3}} \right) ^{2} \left( \kappa_{{2}}-
\kappa_{{3}} \right) ^{2}{\kappa_{{3}}}}\\
&-& 2\, \left( {{\rm e}^{{\frac { \left( \kappa_{{2}}\lambda-1
 \right) \theta}{\lambda}}}}-1 \right)  \left( {{\rm e}^{{\frac {
 \left( \kappa_{{1}}\lambda-1 \right) \theta}{\lambda}}}}-1 \right) 
\frac{ T\,\gamma\,{L_{{0}}}^
{2}{k_{{2}}}^{2}\left( 2+\tau\,\kappa_{{2}}+\tau\,\kappa_{{1}} \right)}{  {m}^{2}{\tau}
^{2} \left( \kappa_{{2}}\lambda-1 \right) \left( \kappa_{{1}}
\lambda-1 \right) \left( \kappa_{{2}}-\kappa_{{3}} \right)
 \left( \kappa_{{1}}-\kappa_{{2}} \right)  \left( \kappa_{{1}}-
\kappa_{{3}} \right) \left( \kappa_{{1}}+\kappa_{{2}} \right)} \\ 
&+& 2\, \left( {{\rm e}^{{\frac { \left( 
\kappa_{{3}}\lambda-1 \right) \theta}{\lambda}}}}-1 \right)  \left( {
{\rm e}^{{\frac { \left( \kappa_{{1}}\lambda-1 \right) \theta}{\lambda
}}}}-1 \right) \frac{ T\,\gamma\,{L_{{0}}}^
{2}{k_{{2}}}^{2}\left( 2+\tau\,\kappa_{{3}}+\tau\,\kappa_{{1}}
 \right)}{  {m}^{2}{\tau}^{2} \left( \kappa_{{3}}\lambda-1 \right)  \left( \kappa_{{1}}\lambda-1 \right)  \left( \kappa_{{2}}-
\kappa_{{3}} \right)  \left( \kappa_{{1}}-\kappa_{{2}} \right)  \left( \kappa_{{1}}-\kappa_{{3}} \right) ^{2} \left( \kappa_{{1}}
+\kappa_{{3}} \right) } \\
&-& 2\,\left( {{\rm e}
^{{\frac { \left( \kappa_{{3}}\lambda-1 \right) \theta}{\lambda}}}}-1
 \right)  \left( {{\rm e}^{{\frac { \left( \kappa_{{2}}\lambda-1
 \right) \theta}{\lambda}}}}-1 \right)  \frac{T\,\gamma\,{L_{{0}}}^
{2}{k_{{2}}}^{2}\left( 2+\tau\,\kappa_{{3}}+
\tau\,\kappa_{{2}} \right)}{{m}^{2}{\tau}^{2} \left( \kappa_{{3}}
\lambda-1 \right) \left( \kappa_{{2}}\lambda-1 \right)
 \left( \kappa_{{1}}-\kappa_{{3}} \right) \left( \kappa_{{1}}-
\kappa_{{2}} \right) \left( \kappa_{{2}}-\kappa_{{3}} \right) ^{2} \left( \kappa_{{2}}+\kappa_{{3}} \right)}
\end{eqnarray*}
\begin{eqnarray*}
\mathcal{A}_2 &=& \left( {{\rm e}^{-{\frac {\theta}{\lambda}
}}}-1 \right) ^{2} \frac{T\,{L_{{0}}}^{2}{k_{{2}}}^{2}}{\left( k_{{1}}+k_{{2}} \right)}
\end{eqnarray*}
\begin{eqnarray*}
\mathcal{A}_3+\mathcal{A}_4 &=&2\,\left( {{\rm e}^{{\frac { \left( \kappa
_{{1}}\lambda-1 \right) \theta}{\lambda}}}}-1 \right)  \left( {{\rm e}
^{-{\frac {\theta}{\lambda}}}}-1 \right) \frac{T\,{L_{{0}}}^{2}{k_{{2}}}^{2}\left( 1+\tau\,\kappa_{{1}}
 \right)}{{m}\,{\tau} \left( \kappa_{{1}}\lambda-1 \right)\left( \kappa_{{1}}-\kappa_{{2}} \right) \left( \kappa_{{1}}-
\kappa_{{3}} \right){\kappa_{{1}}}}\\
&-&2\, \left( {{\rm e}^{{\frac { \left( \kappa_{{2}}\lambda-1 \right) 
\theta}{\lambda}}}}-1 \right)  \left( {{\rm e}^{-{\frac {\theta}{
\lambda}}}}-1 \right)  \frac{T\,{L_{{0}}}^{2}{k_{{2}}}^{2}\left( 1+\tau\,\kappa_{{2}} \right)}{{m}{\tau} \left( \kappa_{{2}}\lambda-1 \right) \left( \kappa_{{
1}}-\kappa_{{2}} \right) \left( \kappa_{{2}}-\kappa_{{3}}
 \right){\kappa_{{2}}}}\\
&+&2\,
 \left( {{\rm e}^{{\frac { \left( \kappa_{{3}}\lambda-1 \right) \theta
}{\lambda}}}}-1 \right)  \left( {{\rm e}^{-{\frac {\theta}{\lambda}}}}
-1 \right)  \frac{T\,{L_{{0}}}^{2}{k_{{2}}}^{2}\left( 1+\tau\,\kappa_{{3}} \right)}{{m}{\tau}\left( \kappa_{{3}}\lambda-1 \right) \left( \kappa_{{1}}-\kappa
_{{3}} \right) \left( \kappa_{{2}}-\kappa_{{3}} \right){
\kappa_{{3}}}}\\
&+&2\,\left( {{\rm e}^{-{
\frac {\theta}{\lambda}}}}-1 \right)^{2} \frac{T\,{L_{{0}}}^{2}{k_{{2}}}^{2}}{{m}\,{\tau}\,{\kappa_{{
1}}}\,{\kappa_{{2}}}\,{\kappa_{{3}}}}\\
\end{eqnarray*}

\begin{eqnarray*}
\mathcal{A}_5 &=& 2\,\left( {{\rm e}^{{\frac { \left( \kappa
_{{1}}\lambda-1 \right) \theta}{\lambda}}}}-1 \right)  \left( {{\rm e}
^{-{\frac {\theta}{\lambda}}}}-1 \right)  \frac{T\,{L_{{0}}}^{2}{k_{{2}}}^{2}\left( 1+\tau\,\kappa_{{1}}
 \right)  \left( k_{{1}}+k_{{2}} \right)}{{m}^{2}{\tau}^{2} \left( 
\kappa_{{1}}\lambda-1 \right){\kappa_{{1}}}^{2} \left( \kappa_{
{1}}-\kappa_{{2}} \right) \left( \kappa_{{1}}-\kappa_{{3}}
 \right){\kappa_{{2}}}{\kappa_{{3}}}}\\
&-&2\,\left( {{\rm e}^{{\frac { \left( \kappa_{{2}}\lambda-1
 \right) \theta}{\lambda}}}}-1 \right)  \left( {{\rm e}^{-{\frac {
\theta}{\lambda}}}}-1 \right)  \frac{T\,{L_{{0}}}^{2}{k_{{2}}}^{2}\left( 1+\tau\,\kappa_{{2}} \right) 
 \left( k_{{1}}+k_{{2}} \right)}{{m}^{2}{\tau}^{2} \left( \kappa_{{2}
}\lambda-1 \right){\kappa_{{2}}}^{2} \left( \kappa_{{1}}-\kappa
_{{2}} \right) \left( \kappa_{{2}}-\kappa_{{3}} \right){
\kappa_{{1}}}{\kappa_{{3}}}}\\
&+&2\,\left( {{\rm e}^{{\frac { \left( \kappa_{{3}}\lambda-1 \right) \theta
}{\lambda}}}}-1 \right)  \left( {{\rm e}^{-{\frac {\theta}{\lambda}}}}
-1 \right)  \frac{T\,{L_{{0}}}^{2}{k_{{2}}}^{2}\left( 1+\tau\,\kappa_{{3}} \right)  \left( k_{{1}}+k_{{2}
} \right)}{{m}^{2}{\tau}^{2} \left( \kappa_{{3}}\lambda-1 \right){\kappa_{{3}}}^{2} \left( \kappa_{{1}}-\kappa_{{3}} \right)
 \left( \kappa_{{2}}-\kappa_{{3}} \right){\kappa_{{1}}}{\kappa_{{2}}}}\\
&+&\left( {{\rm e}^{-{
\frac {\theta}{\lambda}}}}-1 \right) ^{2} \frac{T\,{L_{{0}}}^{2}{k_{{2}}}^{2}\left( k_{{1}}+k_{{2}}
 \right)}{{m}^{2}{\tau}^{2}{\kappa_{{1}}}^{2}{\kappa_{{2}}}^{2}{
\kappa_{{3}}}^{2}}\\
&+& \left( {{\rm e}^{{
\frac { \left( \kappa_{{1}}\lambda-1 \right) \theta}{\lambda}}}}-1
 \right) ^{2} \frac{T\,{L_{{0}}}^{2}{k_{{2}}}^{2}\left( 1+\tau\,\kappa_{{1}} \right) ^{2} \left( m\,{\kappa
_{{1}}}^{2}+k_{{1}}+k_{{2}} \right)}{{m}^{2}{\tau}^{2} \left( \kappa_
{{1}}\lambda-1 \right) ^{2}{\kappa_{{1}}}^{2} \left( \kappa_{{1}}-
\kappa_{{2}} \right) ^{2} \left( \kappa_{{1}}-\kappa_{{3}} \right) ^{
2}}\\
&+&\left( {{\rm e}^{{\frac { \left( \kappa_{{2}}\lambda-1 \right) 
\theta}{\lambda}}}}-1 \right) ^{2} \frac{T\,{L_{{0}}}^{2}{k_{{2}}}^{2}\left( 1+\tau\,\kappa_{{2}}
 \right) ^{2} \left( m\,{\kappa_{{2}}}^{2}+k_{{1}}+k_{{2}} \right)}{{m}^{2}{\tau}^{2} \left( \kappa_{{2}}\lambda-1 \right) ^{2}{\kappa_{{2}}}^{2} \left( \kappa_{{1}}-\kappa_{{2}} \right) ^{2} \left( \kappa_{{2}}-\kappa_{{3}} \right) ^{2}}\\
&+&\left( {{\rm e}^{{\frac { \left( \kappa_{{3}}\lambda-1 \right) 
\theta}{\lambda}}}}-1 \right) ^{2} \frac{T\,{L_{{0}}}^{2}{k_{{2}}}^{2}\left( 1+\tau\,\kappa_{{3}}
 \right) ^{2} \left( m\,{\kappa_{{3}}}^{2}+k_{{1}}+k_{{2}} \right)}{{m}^{2}{\tau}^{2} \left( \kappa_{{3}}\lambda-1 \right) ^{2}{\kappa_{{3}}}^{2} \left( \kappa_{{1}}-\kappa_{{3}} \right) ^{2} \left( \kappa_{{2}}-\kappa_{{3}} \right) ^{2}}\\
&-&2\,\left( {{\rm e}^{{\frac { \left( 
\kappa_{{2}}\lambda-1 \right) \theta}{\lambda}}}}-1 \right)  \left( {
{\rm e}^{{\frac { \left( \kappa_{{1}}\lambda-1 \right) \theta}{\lambda
}}}}-1 \right)  \frac{T\,{L_{{0}}}^{2}{k_{{2}}}^{2}\left( 1+\tau\,\kappa_{{2}} \right)  \left( 1+\tau\,
\kappa_{{1}} \right)  \left( m\,\kappa_{{2}}\,\kappa_{{1}}+k_{{1}}+k_{{2}}
 \right)}{{m}^{2}{\tau}^{2} \left( \kappa_{{2}}\lambda-1 \right) {\kappa_{{2}}}{\kappa_{{1}}} \left( \kappa_{{1}}\lambda-1
 \right) \left( \kappa_{{1}}-\kappa_{{2}} \right) ^{2} \left( 
\kappa_{{2}}-\kappa_{{3}} \right) \left( \kappa_{{1}}-\kappa_{{3
}} \right)}\\
&+&2\,\left( {{\rm e}^{{
\frac { \left( \kappa_{{3}}\lambda-1 \right) \theta}{\lambda}}}}-1
 \right)  \left( {{\rm e}^{{\frac { \left( \kappa_{{1}}\lambda-1
 \right) \theta}{\lambda}}}}-1 \right)  \frac{T\,{L_{{0}}}^{2}{k_{{2}}}^{2}\left( 1+\tau\,\kappa_{{3}}
 \right)  \left( 1+\tau\,\kappa_{{1}} \right)  \left( m\,\kappa_{{3}}\,
\kappa_{{1}}+k_{{1}}+k_{{2}} \right)}{{m}^{2}{\tau}^{2} \left( \kappa
_{{3}}\lambda-1 \right){\kappa_{{3}}}{\kappa_{{1}}}
 \left( \kappa_{{1}}\lambda-1 \right) \left( \kappa_{{1}}-\kappa
_{{3}} \right) ^{2} \left( \kappa_{{2}}-\kappa_{{3}} \right)
 \left( \kappa_{{1}}-\kappa_{{2}} \right)}\\
&-&2\,\left( {
{\rm e}^{{\frac { \left( \kappa_{{3}}\lambda-1 \right) \theta}{\lambda
}}}}-1 \right)  \left( {{\rm e}^{{\frac { \left( \kappa_{{2}}\lambda-1
 \right) \theta}{\lambda}}}}-1 \right)  \frac{T\,{L_{{0}}}^{2}{k_{{2}}}^{2}\left( 1+\tau\,\kappa_{{3}}
 \right)  \left( 1+\tau\,\kappa_{{2}} \right)  \left( m\,\kappa_{{3}}\,
\kappa_{{2}}+k_{{1}}+k_{{2}} \right)}{{m}^{2}{\tau}^{2} \left( \kappa
_{{3}}\lambda-1 \right){\kappa_{{3}}}{\kappa_{{2}}}
 \left( \kappa_{{2}}\lambda-1 \right) \left( \kappa_{{1}}-\kappa
_{{3}} \right) \left( \kappa_{{2}}-\kappa_{{3}} \right) ^{2}
 \left( \kappa_{{1}}-\kappa_{{2}} \right)}.
\end{eqnarray*}

\end{document}